\ifdefined\pdfminorversion
\fi

\documentclass[11pt]{article}

\PassOptionsToPackage{table}{xcolor}
\usepackage{acl}

\usepackage{times}
\usepackage{latexsym}
\usepackage[T1]{fontenc}
\usepackage[utf8]{inputenc}
\usepackage{microtype}
\usepackage{inconsolata}

\usepackage{graphicx}
\usepackage{fancyvrb}
\usepackage{subcaption}
\usepackage{booktabs} 
\usepackage{adjustbox}
\usepackage{tcolorbox}
\tcbuselibrary{breakable}

\usepackage{amsmath}
\usepackage{amssymb}
\usepackage{mathtools}
\usepackage{amsthm}

\usepackage{enumerate}
\newenvironment{itemize*}%
 {\leftmargini=10pt\begin{itemize}%
  \setlength{\itemsep}{0pt}%
  \setlength{\parskip}{7pt}%
  }%
 {\end{itemize}}
\newenvironment{enumerate*}%
 {\begin{enumerate}%
  \setlength{\itemsep}{0pt}%
  \setlength{\parskip}{0pt}}%
 {\end{enumerate}}

\usepackage{tikz}
\usepackage{pgfplots}
\usetikzlibrary{patterns}
\pgfplotsset{compat=1.18}
\usetikzlibrary{patterns}
\usepackage{multirow}
\usepackage{pifont}

\definecolor{ured}{RGB}{220,50,32}
\definecolor{ugreen}{RGB}{34,139,34}

\usepackage[capitalize,noabbrev]{cleveref}

\theoremstyle{plain}

\theoremstyle{definition}

\theoremstyle{remark}

\usepackage[disable,textsize=tiny]{todonotes}

\title{When Scaling Fails: Mitigating Audio Perception Decay of LALMs \\ via Multi-Step Perception-Aware Reasoning}

\author{
  \textbf{Ruixiang Mao}\textsuperscript{1},
  \textbf{Xiangnan Ma}\textsuperscript{1},
  \textbf{Dan Chen}\textsuperscript{1},
  \textbf{Ziming Zhu}\textsuperscript{1},
  \textbf{Yuan Ge}\textsuperscript{1} \\
  \textbf{Haishu Zhao}\textsuperscript{1},
  \textbf{Aokai Hao}\textsuperscript{1},
  \textbf{Yifu Huo}\textsuperscript{1},
  \textbf{Qing Yang}\textsuperscript{1},
  \textbf{Kaiyan Chang}\textsuperscript{1} \\
  \textbf{Xiaoqian Liu}\textsuperscript{1},
  \textbf{Chenglong Wang}\textsuperscript{1},
  \textbf{Weiqiao Shan}\textsuperscript{1},
  \textbf{Qiaozhi He}\textsuperscript{1} \\
  \textbf{Changliang Li}\textsuperscript{3}\thanks{Corresponding authors.},
  \textbf{Zhengtao Yu}\textsuperscript{4},
  \textbf{Tong Xiao}\textsuperscript{1,2}\footnotemark[1],
  \textbf{Jingbo Zhu}\textsuperscript{1,2} \\
  \textsuperscript{1}School of Computer Science and Engineering, Northeastern University, Shenyang, China \\
  \textsuperscript{2}NiuTrans Research, Shenyang, China \quad
  \textsuperscript{3}China Unicom Cloud-Link, Beijing, China \\
  \textsuperscript{4}Kunming University of Science and Technology, China
}

\begin{document}
\maketitle

\begin{abstract}
Test-Time Scaling has shown notable efficacy in addressing complex problems through scaling inference compute. 
However, within Large Audio-Language Models (LALMs), an unintuitive phenomenon exists: 
post-training models for structured reasoning trajectories results in marginal or even negative gains compared to training for direct answering. 
To investigate it, we introduce \textbf{CAFE}, an evaluation framework designed to quantify audio reasoning errors. 
Evaluation results reveal that LALMs encounter a critical bottleneck: \textbf{Audio Perception Decay}, which compromises reasoning performance.
To address it, we propose \textbf{MPAR$^2$}, a strategy that encourages a step-by-step reasoning process with rich perceptual awareness.
With reinforcement learning, MPAR$^2$ improves performance on CAFE from 31.74\% to 63.51\% and effectively mitigates audio perception decay, concurrently enhancing reasoning ability to achieve a strong 74.59\% accuracy on the MMAU benchmark.
Further analysis shows that MPAR$^2$ optimizes audio attention  and enhances reasoning token budget efficiency.
\end{abstract}

\section{Introduction}
Test-Time Scaling~\citep{snell2024scaling} has emerged as a dominant paradigm that enables models to solve complex problems by scaling inference compute, typically realized through extended chains of thought (CoT)~\citep{guo2025deepseek,wei2022chain,brown2024large}.
Inspired by this success, recent works have extended similar strategies to Large Audio-Language Models (LALMs) through reinforcement learning (RL), yielding notable gains on complex audio tasks that demand deep audio understanding and reasoning~\citep{tian2025step,wu2025audio}.
However, a marked exception persists: optimizing models for structured reasoning trajectories via RL results in marginal or even negative gains compared to direct-answering tuning.
As illustrated in Figure~\ref{fig:intro} on the MMAU benchmark~\cite{sakshi2024mmau}, direct-answer RL baselines, which are trained via implicit prompts, outperform CoT-based variants like R1-AQA~\citep{li2025reinforcement} and SARI~\citep{wen2025sari}, where explicit reasoning prompts are utilized to generate \texttt{<thinking>} content. Meanwhile, Audio-Thinker~\citep{wu2025audio} achieves comparable results.
This phenomenon differs from the expectation of scaling that increasing inference compute yields predictable performance improvement.
In light of this, we raise a question:
\textit{what causes the ineffectiveness of extended reasoning in complex audio tasks?}
\definecolor{ugreen}{RGB}{182, 215, 168}
\definecolor{ublue}{RGB}{107, 156, 203}
\definecolor{ured}{RGB}{204, 80, 72}

\definecolor{uyellow}{RGB}{253,186,107}
\definecolor{upurple}{RGB}{175,135,220}
\definecolor{ugray}{RGB}{150,156,165}
\definecolor{ubrown}{RGB}{176,125,086}
\definecolor{uomni}{RGB}{244,165,130}

\definecolor{uone}{RGB}{120,120,120}

\pgfmathsetlengthmacro{\BarW}{15pt}
\pgfmathsetlengthmacro{\BarStep}{18pt}
\pgfmathsetlengthmacro{\GapClusters}{-10pt}

\pgfmathsetlengthmacro{\ShiftLone}{-2.0*\BarStep - 0.5*\GapClusters}
\pgfmathsetlengthmacro{\ShiftLtwo}{-1.0*\BarStep - 0.5*\GapClusters}
\pgfmathsetlengthmacro{\ShiftLthr}{ 0.0*\BarStep - 0.5*\GapClusters}
\pgfmathsetlengthmacro{\ShiftLfou}{ 1.0*\BarStep - 0.5*\GapClusters}
\pgfmathsetlengthmacro{\ShiftLfiv}{ 2.0*\BarStep - 0.5*\GapClusters}

\pgfmathsetlengthmacro{\ShiftRone}{-1.0*\BarStep + 0.5*\GapClusters}
\pgfmathsetlengthmacro{\ShiftRtwo}{ 0.0*\BarStep + 0.5*\GapClusters}
\pgfmathsetlengthmacro{\ShiftRthr}{ 1.0*\BarStep + 0.5*\GapClusters}

\tikzset{
  hatchbar/.style={
    postaction={draw=none, 
    pattern=north east lines,
    }
  },
  hatchCross/.style={
    postaction={draw=none, pattern=crosshatch}
  }
}

\begin{figure}[t]
\centering
\begin{tikzpicture}

\begin{axis}[
  ybar,
  bar width=\BarW,
  width=1.1\linewidth,
  height=0.70\linewidth,
  ymin=50, ymax=73,
  ytick={50, 55, 60, 65},
  symbolic x coords={R1-AQA, SARI, Audio-Thinker, ~~Audio-Thinker$^*$},
  xtick={R1-AQA, SARI, Audio-Thinker,~~Audio-Thinker$^*$},
  enlarge x limits=0.18,
  axis line style={thick},
  tick style={thick,draw=none},
  label style={font=\normalsize},
  yticklabel style={font=\scriptsize},
  xticklabel style={
  font=\scriptsize,
  yshift=0pt,            
  text height=0.4ex,     
  text depth=0.1ex,      
  },
  ymajorgrids,
  grid style={dashed, gray!30},
  legend style={
    font=\scriptsize,
    yshift=-0.9cm,
    draw=none,
    fill=none,
    text opacity=1,
    rounded corners,
    at={(axis description cs:0.24,1.23)}, 
    anchor=north,     
    legend cell align=left,
    nodes={anchor=base},
    /tikz/every node/.append style={yshift=-1pt}
  },
  legend image code/.code={
    \path[#1, draw=none, fill opacity=0.95] (0pt,-1pt) rectangle (8pt,7pt);
  },
  legend columns=3,
  clip=false,
  nodes near coords={
  \textbf{\pgfmathprintnumber[fixed,precision=1]{\pgfplotspointmeta}}
},
  nodes near coords align={vertical},
  every node near coord/.append style={
    font=\scriptsize,
    text=black,
    yshift=0.5pt
  }
]

\addplot+[
  forget plot,
  draw=black, 
  pattern color=ublue,
  fill=ublue!45,
  bar shift=\ShiftLtwo
] coordinates {(R1-AQA,64.50)};

\addplot+[
  forget plot,
  draw=black, 
  fill=ured!45,
  pattern color=ured,
  bar shift=\ShiftLthr
] coordinates {(R1-AQA,61.10)};

\addplot+[
  forget plot,
  draw=black,  fill=ublue!45,
  bar shift=\ShiftLtwo
] coordinates {(SARI,59.90)};

\addplot+[
  forget plot,
    draw=black,  fill=ured!45,
  bar shift=\ShiftLthr
] coordinates {(SARI, 59.5)};

\addplot+[
  forget plot,
    draw=black,  fill=ublue!45,
  bar shift=\ShiftLtwo
] coordinates {(Audio-Thinker, 63.80)};

\addplot+[
  forget plot,
    draw=black,  fill=ured!45,
  bar shift=\ShiftLthr
] coordinates {(Audio-Thinker, 65.00)};

\addplot+[
  forget plot,
    draw=black,  fill=ublue!45,
  bar shift=\ShiftLtwo
] coordinates {(~~Audio-Thinker$^*$,69.70)};

\addplot+[
  forget plot,
    draw=black,  fill=ured!45,
  bar shift=\ShiftLthr
] coordinates {(~~Audio-Thinker$^*$,69.80)};

\pgfplotsset{
    bar legend/.style={
        legend image code/.code={
            \draw[#1] (0cm,-0.12cm) rectangle (1cm,0.18cm);
        }
    }
}

\addlegendimage{bar legend={
      draw=black, 
    fill=ublue!45,sharp corners,
    fill opacity=0.9
}}
\addlegendentry{RL}

\addlegendimage{bar legend={
      draw=black,  
    fill=ured!45,sharp corners,
    fill opacity=0.9
}}
\addlegendentry{RL$^\dagger$}

\node[
  font=\small,
  text=red!60!black
] at (rel axis cs:0.46,0.8)
{\textbf{Negative and Marginal Gains}};

\draw[thick](rel axis cs:0,0) -- (rel axis cs:1,0);

\end{axis}
\end{tikzpicture}

\caption{Comparison of direct-answer RL training and RL with explicit reasoning prompts (denoted by RL$^\dagger$).
Three RL variants (R1-AQA, Audio-Thinker, and SARI) are tested on the MMAU benchmark. Here, * denotes Qwen2.5-Omni as the base model; otherwise, Qwen2-Audio is used. More details of the prompts used and response cases are presented in Appendix~\ref{sec:prompts_rl}.
}
\label{fig:intro}
\vspace{-10pt}
\end{figure}

Recent research has attempted to tackle similar challenges. Step-Audio-R1~\citep{tian2025step} attributes this issue to an over-reliance on textual reasoning, exploring modality-grounded frameworks to mitigate the gap.
Meanwhile, an analysis in MMAR~\cite{ma2025mmar} reveals audio event perception and utilization errors are the dominant factor.
However, despite these efforts, a systematic empirical investigation of this problem is still rare.

To fill this void, we introduce \textbf{C}omprehensive \textbf{A}udio \textbf{F}idelity \textbf{E}valuation framework (\textbf{CAFE}), which is designed to quantify audio event fidelity during reasoning process. 
Leveraging the “LLM-as-a-Judge” paradigm~\cite{zheng2023judging}, CAFE extracts and counts audio events from reasoning processes to quantify perception and utilization errors.
Experimental results reveal that current audio models lack the ability to stably anchor acoustic cues across long generated sequences, \textit{audio perception decay} prevents extended reasoning from delivering its intended positive impact. 
As the reasoning chain extends, the model gradually omits or hallucinates events, with the accumulation of such errors degrading the accuracy of logical deductions.

To address this limitation, we propose \textbf{MPAR$^2$}, a
\textbf{M}ulti-step \textbf{P}erception-\textbf{A}ware \textbf{R}easoning and \textbf{R}eview strategy designed to mitigate perception decay, thereby enhancing reasoning capability.
Extending the power of CoT, MPAR$^2$ establishes a “think-while-listening” paradigm. 
It follows a structured pipeline: initially enforcing a fine-grained scan for explicit audio perception, followed by step-wise perception-rich question decomposition, and concluding with a review process to check overall reasoning details and rationality. 
To instantiate this strategy, we adopt a two-stage training scheme. First, supervised cold-start training endues formatted reasoning ability. 
In the second stage, RL further optimizes the model for audio perception and reasoning. Experiments show MPAR$^2$ achieves 74.59\% on MMAU (origin) test-mini and 60.32\% on MMAR. 
Under the CAFE evaluation, MPAR$^2$ achieves leading perception and utilization accuracy. 
Further analysis shows that MPAR$^2$ reinforces
audio attention during reasoning and dynamically adjusts reasoning budget based on question complexity. 
In summary, our contributions are:
\vspace{-3pt}
\begin{itemize*}
    \item 
We introduce \textbf{CAFE}, an evaluation framework that provides a powerful quantitative tool for assessing perception and reasoning quality during the reasoning process of audio models.
\vspace{-2pt}
    \item 
We present the first systematic analysis of \textit{audio perception decay} in LALMs, where longer reasoning chains lead to degraded audio perception, resulting in marginal or even negative gains from complex CoT reasoning.
\vspace{-2pt}
     \item 
We propose \textbf{MPAR$^2$}, an RL-based strategy designed to enhance both audio event perception and utilization. Experiments show that MPAR$^2$ achieves strong perception and reasoning accuracy, while reinforcing audio attention and enhancing reasoning token budget efficiency.
\end{itemize*}

\section{Related Work}

\paragraph{Large Audio-Language Models}
As large language models (LLMs) continue to advance, models that integrate multiple modalities have gradually gained growing focus. These multimodal LLMs are capable of jointly modeling and reasoning over various inputs, including audio. Recent LALMs, such as Qwen2-Audio~\cite{chu2024qwen2}, Qwen2.5-Omni~\cite{Xu2025Qwen25OmniTR}, DeSTA2.5-Audio~\citep{lu2026desta2}, and MiMo-Audio~\cite{zhang2025mimo}, have demonstrated notable effectiveness in understanding and processing acoustic information.

\paragraph{Large Audio Reasoning Models}
\vspace{-4pt}
Recent research has increasingly focused on enhancing LALMs with reasoning capability. While early efforts like Audio-CoT~\citep{ma2025audio} and Audio-Reasoner~\citep{xie2025audio} introduced CoT via prompting or supervised fine-tuning (SFT), but achieved limited gains and failed to unlock the reasoning potential of audio models. Subsequently, reinforcement learning has recently pushed boundaries. Methods such as R1-AQA~\citep{li2025reinforcement} and SARI~\citep{wen2025sari} utilize RL to optimize structured CoT reasoning, whereas Omni-R1~\citep{rouditchenko2025omni} adopts a direct-answer strategy that surprisingly surpasses CoT. However, results on the variants of R1-AQA and SARI reveal an unintuitive behavior in audio test-time scaling. 
Although other strategies like Audio-Thinker~\citep{wu2025audio} explore an adaptive thinking mode, Step-Audio-R1~\citep{tian2025step} applies modality-grounded frameworks to further improve reasoning capability, and CESAR~\citep{fan2025incentivizing} first analyses it and applies fine-grained reward method.
\paragraph{Adaptive Reasoning and Difficulty Awareness}
The success of reasoning models such as OpenAI o1~\cite{jaech2024openai} demonstrates the effectiveness of test-time scaling, while also highlighting the practical need to reduce computational redundancy. Prior work shows that LLMs implicitly encode problem difficulty in their hidden representations~\citep{lee2025probing}, enabling adaptive reasoning termination. Building on this insight, S-GRPO~\citep{dai2025s} employs reinforcement learning with decaying rewards to encourage early correct exits, while data-centric methods~\citep{waheed2025less} further improve token efficiency by distilling difficulty-aware reasoning length through compressed CoT trajectories. 

\section{Reasoning-Time Audio Event Perception and Utilization Probing}
\label{sec:cafe}

In this section, we define the task of reasoning-time audio event perception and  utilization probing and introduce the CAFE framework. Subsequently, we present and analyze evaluation results across LALMs, Large Audio Reasoning Models (LARMs), and commercial models.

\paragraph{Task Formulation}
Given ground-truth audio content and Question-Answer (QA) pairs, the reasoning-time audio event perception and  utilization probing task aims to quantify the identification and utilization of audio events during model reasoning, employing metrics such as perception accuracy, utilization accuracy, and omission rate. 
We leverage a powerful textual LLM to process the probing inputs. 
This approach enables robust generalization across semantic variations, while ensuring the precise extraction of audio events strictly required for the question-response context.
Furthermore, it enables the quantification of audio fidelity throughout statistically analyzing the extracted audio events.

\begin{figure}[t]
    \centering
    \scalebox{0.8}{\includegraphics[width=\columnwidth]{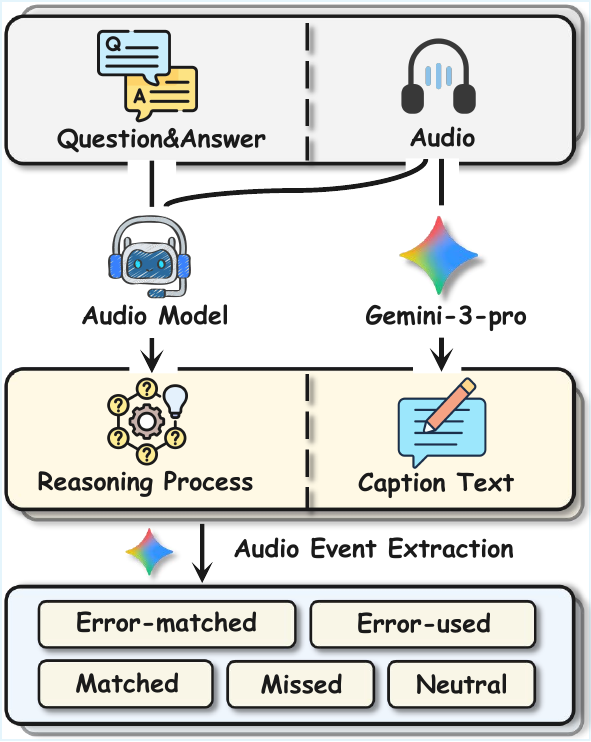}}
    \caption{Overview of the CAFE framework evaluation pipeline. First, audio model generates a reasoning process based on the audio-QA input. Meanwhile, the audio is captioned by Gemini-3-Pro. Finally, Gemini-3-Pro utilizes the inputs, the reasoning process, and the captions to extract different audio events.}
    \label{fig:overview}
    \vspace{-8pt}
\end{figure}

\subsection{Overview}
\label{sec:CAFE_overiew}
To address this task, we introduce the CAFE framework.
Formally, employing the probing textual model $\mathcal{P}$, the audio event extraction process can be formulated as:
\begin{equation}
    \mathcal{E} = \mathcal{P}(O, Q, A, C)
\end{equation}
where $O$ represents the model's reasoning output generated by the model, $Q$ and $A$ denote the input question and the target answer, and $C$ is the ground truth audio caption. 
The same abbreviations are used throughout the subsequent formulas.
The resulting output $\mathcal{E}$ consists of categorized audio events with labels $E_i$, where aspect $E_i\in \{\textit{Matched},\allowbreak \textit{Hallucinated},\allowbreak \textit{Misused},\allowbreak \textit{Missed},\allowbreak \textit{Neutral}\}$, defined as follows: (1) \textit{Matched}: existing audio events correctly identified and utilized during the reasoning process to address the question. (2) \textit{Hallucinated}: events that are hallucinatory or misidentified during reasoning; (3) \textit{Misused}: existing events that are logically misused, leading to incorrect conclusions; (4) \textit{Missed}: existing events required for the answer but omitted during reasoning; and (5) \textit{Neutral}: existing events mentioned but irrelevant to the reasoning outcome. Furthermore, we extract and count these categorized events to derive specific metrics below.

\input{figures/fig2}

\paragraph{Metric Formulation}
\vspace{-4pt}
We define four distinct metrics based on the extracted event categories. Let $N_{(\cdot)}$ denote the count of events for each category. We first define the prediction space $N_{\textit{pred}}$ and the target space $N_{\textit{tgt}}$ as follows:
\begin{equation}
\qquad\quad
\begin{aligned}
N_{\text{pred}} &= N_{\text{mat}} + N_{\text{hal}} + N_{\text{misuse}} + N_{\text{neu}}, \\
N_{\text{tgt}}  &= N_{\text{mat}} + N_{\text{miss}}
\end{aligned}
\vspace{-3pt}
\end{equation}
where $N_{\textit{pred}}$ denotes the total number of audio events in model reasoning process, and $N_{\textit{tgt}}$ represents the complete set of ground-truth events required to answer the question. Based on these, we derive the following metrics:
\vspace{-4pt}
\begin{itemize}
    \item \textit{Perception Accuracy} ($\textit{Acc}_{\text{per}}$) is defined as the ratio of correctly identified events related to the QA task, formulated as $\textit{Acc}_{\text{per}} = \frac{N_{\text{mat}}}{N_{\text{pred}}}$. $\textit{Acc}_{\text{per}}$ is aimed at perception in  reasoning rather than simple audio caption.
    \vspace{-5pt}
    \item \textit{Perception Error Rate} ($\textit{Err}_{\text{per}}$) quantifies the rate of hallucinatory or misidentified events via $\textit{Err}_{\text{per}} = \frac{N_{\text{hal}}}{N_{\text{pred}}}$.
    \vspace{-5pt}
    \item \textit{Usage Error Rate} ($\textit{Err}_{\text{use}}$) assesses events that are correctly perceived but logically misused, which can be formulated as $\textit{Err}_{\text{use}} = \frac{N_{\text{misuse}}}{N_{\text{pred}}}$.
    \vspace{-3pt}
    \item \textit{Omission Error Rate} ($\textit{Err}_{\text{omit}}$) quantifies the events required for the answer that the model failed to capture, defined as $\textit{Err}_{\text{omit}} = \frac{N_{\text{miss}}}{N_{\text{tgt}}}$.
\end{itemize}
In particular, $\textit{Acc}_{\text{per}}$ offers a holistic assessment of the audio model's perceptual fidelity during reasoning. This metric serves as the quantitative basis for Figure \ref{fig:six_plots}, where we track perceptual performance as reasoning length increases, ultimately revealing the phenomenon of audio perception decay.

\paragraph{Framework details}
Within CAFE, Gemini-3-Pro generates captions for the benchmark audio and serves as the probing model $\mathcal{P}$. Replacing it with GPT-5 and using semantically equivalent prompts in an ablation study leads to performance variations within ±2.1\%. The overall workflow of the framework is shown in Figure \ref{fig:overview}. Additional prompts and details are provided in Appendix \ref{sec:app_A}.

\subsection{Evaluation}

\paragraph{Experimental Setup}

For models that lack explicit reasoning output, we employ designed prompts to elicit reasoning. 
A unified prompt is used for the remaining models. Details and additional analysis are presented in Appendix \ref{sec:app_B}.

\subsection{Evaluation Results}

To analyze the relationship between reasoning length and both perception and reasoning accuracy, we selected two representative models from each category: LALMs, LARMs, and commercial models, as shown in Figure~\ref{fig:six_plots}. Table~\ref{table:cafe_results} presents the overall results, including the MPAR$^2$ method, which is detailed in Section \ref{sec:method}. Based on these results, we conduct analysis and identify two findings.

\paragraph{Audio Perception Decay Across Extended Reasoning}
As illustrated in Figure~\ref{fig:six_plots} (a)-(d), open-source LARMs and LALMs exhibit a clear downward trend: both perception and reasoning accuracy decline as reasoning length increases, eventually dropping to zero.
Pearson~\cite{benesty2009pearson} analysis confirms a robust positive correlation between these two metrics across (a)-(d) groups, yielding coefficients ranging from 0.65 to 0.91, where the majority reach statistical significance with $p < 0.05$.
In contrast, commercial models (e)-(f) maintain stable and superior performance without exhibiting decay trend. 
Thus, the observed variations in these metrics show that perception struggles to maintain a robust connection with the expanding reasoning chain. Instead, perception accuracy degrades as reasoning length increases, which we term \textbf{“Audio Perception Decay”}, leading to an accumulation of errors that drives the degradation of reasoning capability.

\paragraph{Weak Perception during Reasoning} 
As shown in Table~\ref{table:cafe_results}, most audio models perform poorly on CAFE for MMAR. Notably, open-source models rarely surpass the threshold of 40\% $\text{Acc}_{\textit{per}}$, and even advanced systems like Gemini-2.5-Flash reach only 54.81\%, with the majority falling below 30\%. 
However, curated RL strategies prove to significantly enhance capabilities.
Specifically, Omni-R1 (51.21\%) substantially outperforms its Qwen2.5-Omni baseline (31.74\%), while Step-Audio-R1.1 (56.18\%) and Step-Audio-R1 (53.08\%) similarly achieve remarkable performance.
For instance, the Step-Audio-R1 series exhibits no clear bottleneck across the four metrics, whereas other models show marked disparities, underscoring that reasoning capability is critically sensitive to even partial perceptual deficits.
These findings reveal that \textit{effective audio reasoning requires a strong grounding in audio perception}, rather than merely producing well-formatted but semantically disconnected outputs.

\begin{table}
\centering
\resizebox{\linewidth}{!}{
\setlength{\tabcolsep}{5pt}
\begin{tabular}{@{}lccccc@{}}   
\toprule
\textbf{Model} 
&  \textbf{$\textit{Acc}_{\text{per}}$~$\uparrow$}  
&  \textbf{$\textit{Err}_{\text{per}}$~$\downarrow$}
& \textbf{$\textit{Err}_{\text{use}}$~$\downarrow$}
&  \textbf{$\textit{Err}_{\text{omit}}$~$\downarrow$}
& \textbf{$\textit{MMAR}_{\text{acc}}$~$\uparrow$}\\
\midrule
\addlinespace[0.5mm] 

\rowcolor{gray!20}
\multicolumn{6}{c}{\textbf{\textit{Large Audio Language Models}}} \\

MiMo-Audio & 42.79  & 27.70 & 22.40 & 48.84 & 59.87 \\
Qwen2.5-Omni-7B & 31.74  & 29.61 & 28.69 & 57.37 & 55.20 \\
DeSTA2.5-Audio$^\dagger$ & 23.19  & 33.95 & 35.57 & 69.32 & 41.60 \\
Phi-4-Multimodal & 20.43  & 36.75 & 30.32 & 73.41 & 39.80 \\
Qwen2-Audio-Instruct & 8.27  & 56.46 & 28.58 & 81.01 & 29.90 \\
Omni-R1 & 51.21  & 28.79 & 12.90 & 42.65 & 62.10 \\

\rowcolor{gray!20}
\multicolumn{6}{c}{\textbf{\textit{Large Audio Reasoning Models}}} \\
Step-Audio-R1.1  & \underline{56.18} & 23.77 & \underline{11.25} & \underline{21.49} & \textbf{67.50}\\
Step-Audio-R1  & 53.08  & 24.82 & 12.42 & \textbf{6.77} & \underline{67.40} \\
Audio-Flamingo-3$^\dagger$ & 41.39  & 26.03 & 23.18 & 52.87 & 56.40 \\
Audio Reasoner & 27.44  & 25.55 & 32.58 & 55.72 & 36.80 \\

\rowcolor{gray!20}
\multicolumn{6}{c}{\textbf{\textit{Commercial Models}}} \\

Gemini-2.5-Flash & 54.81  & \underline{19.27} & 15.96 & 33.24 & 66.30 \\
GPT-4o-Audio & 48.68  & \textbf{17.20} & 21.71 & 38.05 & 63.80 \\
\midrule
\textbf{MPAR$^2$-7B} & \textbf{63.51}  & 23.14 & \textbf{7.74} & 30.59 & 60.32 \\
\bottomrule
\end{tabular}
}
\caption{Results of various audio models on the MMAR benchmark within the CAFE framework. $^\dagger$ denotes $\textit{MMAR}_{\text{acc}}$ evaluated by GPT-5 judge. All results are averaged over three inference runs. The best performance in each group is in \textbf{bold} and the second best one is \underline{underlined}. The remaining table settings are the same.}
\label{table:cafe_results}
\vspace{-10pt}
\end{table}

\begin{figure*}[!th]
    \centering
    \includegraphics[width=\textwidth]{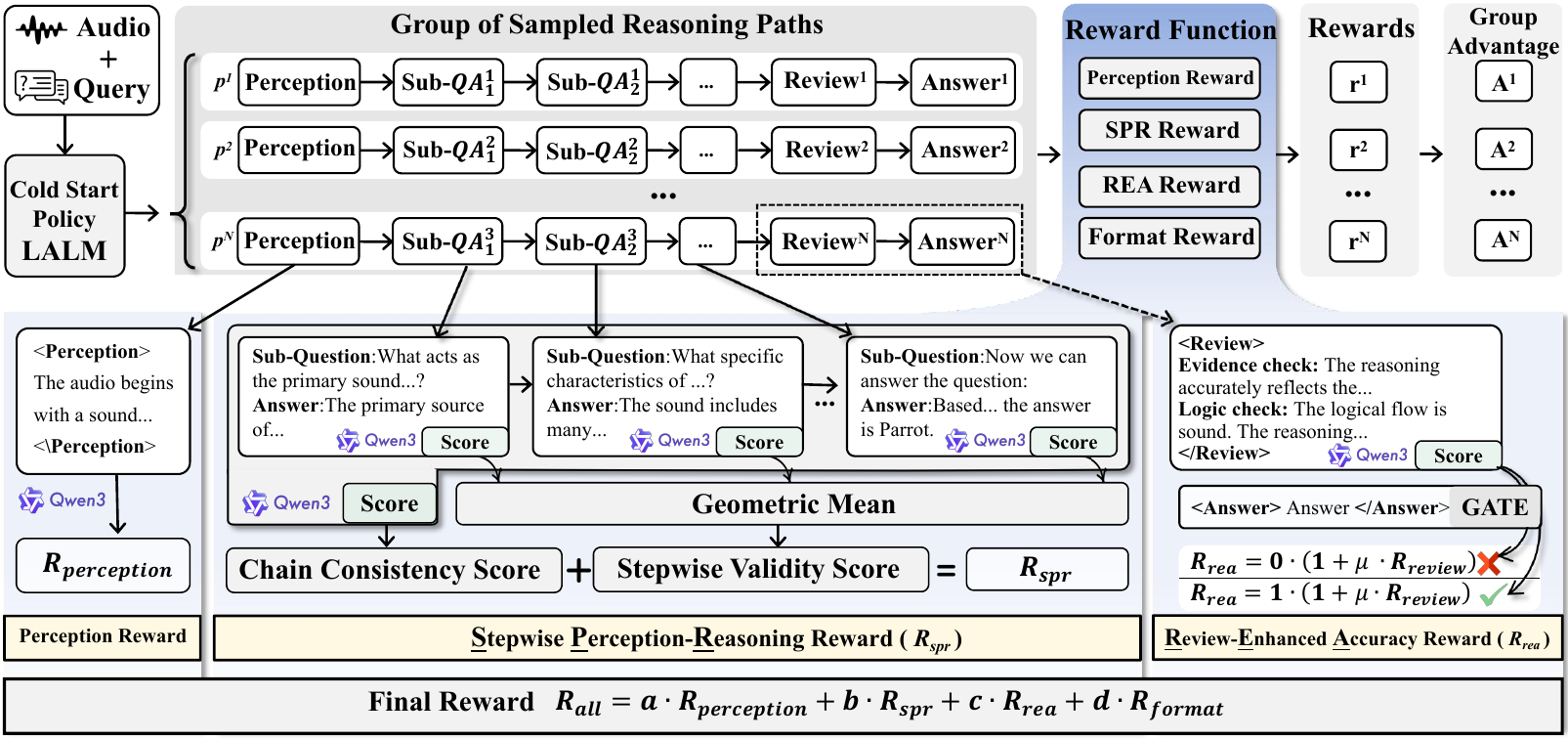}    
    \caption{
     An illustration of MPAR$^2$ training pipeline. The top of the figure outlines the overall GRPO training setup. The lower section provides a view of the reward design. Cold start policy model is derived from SFT-trained model.
    }
    \vspace{-10pt}
    \label{fig:main_image}
\end{figure*}

\section{Methodology}
\label{sec:method}

To address the perception decay of reasoning over extended sequences, we propose a novel reasoning paradigm, \textbf{M}ulti-step \textbf{P}erception \textbf{A}ware \textbf{R}easoning and \textbf{R}eview strategy (\textbf{MPAR$^2$}), which is designed to structurally reinforce audio perception and reasoning while maintaining compatibility with inference-time scaling strategies. In this section, we present the MPAR$^2$ CoT structure, reward design and describe its two-stage training procedure. The overview of MPAR$^2$ is in Figure \ref{fig:main_image}.

\subsection{CoT structure of \texorpdfstring{MPAR$^2$}{MPAR$^2$} Framework}
We begin by defining the reasoning process under the MPAR$^2$ paradigm. Each reasoning chain follows three stages: (1) \textit{Perception}: Explicitly describe the audio events related to the QA task.
Detail their inter-relationships, strictly following the time sequence; (2) \textit{Reasoning}: Dynamically decompose the original question into a structured sequence of sub-questions, adjusting the sub-question steps based on the question’s difficulty. Then, solve the sub-questions using audio-perception details and reasoning, until the core inquiry is fully resolved. (3) \textit{Review}: Double-check the fidelity of the audio events used during reasoning, and ensure the logical consistency. Make necessary corrections to the final answer if errors are detected.

\subsection{Multi Rewards Designs}

To both refine perception quality and promote adaptive reasoning budgets within the MPAR$^2$ paradigm, we leverage Group Relative Policy Optimization (GRPO)~\cite{shao2024deepseekmath}. This optimization process is steered by three generative rewards and one format reward, which jointly work to enhance both reasoning and perceptual fidelity. Specifically, we utilize a powerful textual LLM $S$ as a score model to obtain process-level rewards, guided by designed prompts.
The prompt designs and more details are provided in Appendix \ref{sec:app_D}.

\paragraph{Perception Reward}
\vspace{-5pt}
Given the severe perceptual limitations and audio perception decay during reasoning, MPAR$^2$ begins the reasoning process with an explicit audio perception step. 
It incorporates a perception reward to ensure high-precision estimation, as shown in Figure \ref{fig:main_image}. Formally, using the scoring model $S$, the reward is defined as:
\begin{equation}
R_\text{perception} = S(O_{\text{perception}},Q,A,C)
\end{equation}
where $O_{\text{perception}}$ represents the perception part of CoT . Specifically, given the ground truth caption with QA, we prompt Qwen3-32B as the model $S$ to assess the generated perception.
The model outputs a score $R_{\text{perception}} \in [0,1]$ as reward with a specific focus on the audio events essential for solving the problem, evaluating them based on faithfulness, completeness, and consistency.

\paragraph{Stepwise Perception-Reasoning Reward}
\vspace{-4pt}
Prior studies~\cite{marjanovic2025deepseek,lee2025evaluating} demonstrate that structured reasoning trajectories, such as problem decomposition, significantly enhance complex problem-solving. 
Meanwhile, existing approaches such as Audio-Thinker~\cite{wu2025audio} primarily rewards the overall reasoning process in a coarse manner, but lacks fine-grained guidance over reasoning quality and explicit emphasis on audio perception.
Inspired by these findings, we introduce Stepwise Perception-Reasoning Reward (SPR Reward) to reinforce structured reasoning.

SPR Reward uses two key criteria: \textit{stepwise validity} and \textit{chain consistency}. Specifically, stepwise validity evaluates the event completeness and logical soundness of each sub-QA. It aims for high efficiency and direct resolution within the current step, ensuring that every question posed is both critical and perception-rich. 
We calculate these step scores using the geometric mean~\cite{mcalister1879xiii}, which is described in Eq. \ref{eq:gm}, a method chosen to enforce uniform quality across the entire process by strictly penalizing any single step failure. 

Moving beyond individual steps, chain consistency evaluates the logic of the entire reasoning path. 
Crucially, this mechanism encourages the model to adaptively adjust its sub-question decomposition based on question difficulty, thereby controlling the reasoning budget. 
It requires making effective use of detailed audio descriptions and analysis to establish a clear reasoning, while enforcing a strict order in which audio analysis precedes and supports final problem-solving.

\begin{table*}[t!]
\centering
\tiny 
\setlength{\defaultaddspace}{1pt}
\renewcommand{\arraystretch}{0.9}
\resizebox{\linewidth}{!}{
\begin{tabular}{@{}lccccccccc@{}}
\toprule
\multirow{3}{*}{\textbf{Model}} & 
\multirow{3}{*}{\textbf{Size}} & 
\multicolumn{4}{c}{\textbf{MMAU (Test-mini)}} & 
\multicolumn{4}{c}{\textbf{MMAR}} 
\\ \cmidrule(lr){3-6} \cmidrule(lr){7-10}

& & Sound$\uparrow$ & Music$\uparrow$ & Speech$\uparrow$ & Avg$\uparrow$ & Sound$\uparrow$ & Music$\uparrow$ &  Speech$\uparrow$ & Avg$\uparrow$ \\ 
\midrule

\rowcolor{gray!20}
\multicolumn{10}{c}{\textbf{\textit{Large
Audio Language Models}}} \\

MiMo-Audio & 7B & \textbf{80.78} & 70.96 & 63.06 & 71.60 & 58.79 & 43.69 & 60.54 & 59.87 \\
Qwen2.5-Omni & 7B &  69.40 & 66.80 & 61.60 & 65.90 & 53.94 & 42.23 & 58.84 & 55.20 \\
DeSTA2.5-Audio$^\dagger$ & 8B  & 58.26 & 44.61 & 58.26 & 53.70 & 53.74 & 24.76 & 29.70 & 41.60 \\
Phi-4-Multimodal & 5.6B & 57.96 & 51.80 & 48.35 & 52.70 & 32.12 & 30.58 & 45.58 & 39.80 \\
Qwen2-Audio-Instruct & 8.4B  & 49.85 & 52.40 & 42.94 & 48.40 & 33.33 & 24.27 & 32.31 & 29.90 \\
Omni-R1 & 8.4B  & 73.60 & \textbf{74.30} & 66.10 & 71.30 & \underline{67.30} & \underline{51.50} & 64.30 & 62.10 \\
Kimi-Audio & 11.3B & 61.68 & \underline{73.27} & 60.66 & 65.00 & 48.48 & 29.13 & 42.52 & 40.60 \\

\rowcolor{gray!20}
\multicolumn{10}{c}{\textbf{\textit{Large
Audio Reasoning Models}}} \\

Audio-Flamingo-3$^\dagger$ & 7B & \underline{79.58} & 72.46 & 60.36 & 70.80 & 58.26 & 44.61 & 58.26 & 56.40 \\
Audio Reasoner$^*$ & 8.4B & 60.06 & 64.30 & 60.70 & 61.71 & 43.64 & 33.50 & 32.99 & 36.80 \\
SARI$^*$ & 7B & 72.75 & 67.22 & 61.26 & 67.08 & - & - & - & - \\
R1-AQA$^*$ & 8.4B & 68.77 & 64.37 & 63.66 & 65.60 & - & - & - & - \\
Audio-Thinker$^*$ & 7B & 77.48 & 70.36 & 73.37 & 73.70 & \textbf{68.32} & \textbf{53.88} & 64.29 & 65.30 \\
Step-Audio-R1 & 33B & 72.97 & 61.68 & \underline{74.47} & 69.70 & 57.58 & 45.63 & \underline{77.89} & \underline{67.40} \\
Step-Audio-R1.1 & 33B & 72.67 & 72.46 & \textbf{77.18} & \underline{74.10} & 60.00 & 45.15 & \textbf{79.25} & \textbf{67.50} \\

\rowcolor{gray!20}
\multicolumn{10}{c}{\textbf{\textit{Commercial Models}}} \\

Gemini-2.5-Flash & - & 67.56 & 61.48 & 63.76 & 64.27 & 53.33 & 50.97 & 75.85 & 66.30 \\
GPT-4o-Audio & - & 60.36 & 57.19 & 67.27 & 61.60 & 54.55 & 48.54 & 69.39 & 63.80 \\

\midrule
\textbf{MPAR$^2$-7B} & 7B & 77.98 & 71.30 & \underline{74.47} & \textbf{74.59} & 62.40 & 44.30 & 64.26 & 60.32 \\
\quad w/o $R_\text{perception}$ & 7B & 75.84 & 72.90 & 71.34 & 73.36 & 60.10 & 42.60 & 63.82 & 58.84 \\
\quad w/o $R_{\text{spr}}$ & 7B & 77.74 & 70.68 & 73.78 & 74.08 & 61.90 & 43.80 & 63.16 & 59.62 \\
\quad w/o $R_{\text{spr}}$ and $R_\text{perception}$ & 7B & 77.19 & 70.66 & 71.43 & 73.11 & 59.60 & 42.40 & 62.60 & 58.20 \\
\quad w/o $R_{\text{spr}}$, $R_\text{perception}$ and $R_\text{rea}$$^\ddagger$  & 7B & 77.54 & 67.32 & 73.27 & 72.81 & 58.90 & 41.80 & 61.86 & 57.52 \\

\bottomrule
\end{tabular}
}
\caption{Accuracy (\%) on MMAU (original) and MMAR benchmark. $^\dagger$ denotes results evaluated by GPT-5. * means results are reproduced, while all other results are averaged over three inference runs. $^\ddagger$ means the ablation setting with a simple accuracy reward.}
\label{table:benchmark_res}
\vspace{-10pt}
\end{table*}

With these criteria, we define SPR Reward as:
\begin{equation}
\qquad
R_{\text{spr}}
= \theta \cdot \bar{S}_{\text{sub-reason}}
+ (1-\theta)\cdot {S}_{\text{all-reason}}
\end{equation}
$\bar{S}_{\text{sub-reason}}$ and ${S}_{\text{all-reason}}$  are calculated by:
\begin{align}\quad
\bar{S}_{\text{sub-reason}}
&= \left( \prod_{i=1}^{n}
S\big(O_{\text{sub-reason}}^{(i)}, Q, A, C\big)
\right)^{\frac{1}{n}} \label{eq:gm}\\
S_{\text{all-reason}}
&= S\big(O_{\text{all-reason}}, Q, A, C\big)
\end{align}
where $\theta\in [0,1]$ allocates the reward weights between individual reasoning steps and entire reasoning path, which is set to 0.7. $O_{\text{sub-reason}}^{(i)}$ denotes the $i$-th sub-reasoning step, and $O_{\text{all-reason}}$ is the overall reasoning content. 
Notably, for stepwise validity and chain consistency evaluation, in addition to the reasoning content, we also input the audio caption $C$, the question $Q$, and the ground-truth answer $A$, thereby improving the evaluation quality.
 
\paragraph{Review-Enhanced Accuracy Reward} 
Current explicit reasoning in audio models is inherently one-pass. Once the model generates reasoning steps, it continues without looking back to re-examine the alignment between the audio content and its text. Consequently, if an initial perceptual error occurs (e.g., misidentifying a “cat” as a “dog”), subsequent reasoning steps merely reinforce this error rather than correcting it. This lack of an “auditory feedback loop” necessitates a mechanism for backward verification. To address this, we introduce the Review-Enhanced Accuracy Reward (REA Reward), which integrates this double-check with a direct reward for the correctness of the final answer. We calculate it:
\begin{equation}
R_\text{rea} = R_\text{acc} \cdot (1+ \mu \cdot R_\text{review})
\label{eq:rear}
\end{equation}
where we set $\mu$ to 0.5, and $R_\text{review}$ is defined as:
\begin{equation}
\begin{aligned}
R_\text{review} = S(&O_{\text{review}}, O_{\text{all-reason}}, Q, A, C)
\end{aligned}
\end{equation}
where $O_\text{review}$ denotes the review part of model output. $R_{\text{acc}}$ is defined as a binary reward, taking the value of 1 if the generated answer aligns with the ground truth, and 0 otherwise. 
Notably, we treat $R_{\text{acc}}$ as a gating factor to prevent that review process is rewarded but the answer is still wrong, as shown in Eq. \ref{eq:rear}. This ensures that the review reward is only granted when the answer is correct.

\paragraph{Format Reward} 
We define a format reward $R_{\text{format}}$ of 1 for outputs complying with the MPAR$^2$ structure illustrated in Figure \ref{fig:main_image}, and 0 otherwise.

\paragraph{Overall Reward} 
After assigning weights to each component, the final reward $R_{\text{all}}$ is defined as:
\begin{equation}
R_{\text{all}} = \alpha \cdot R_{\text{perception}} + \beta \cdot R_{\text{spr}} + \gamma \cdot R_{\text{rea}} + \delta \cdot R_{\text{format}}
\end{equation}
where we set $\alpha$, $\beta$, $\gamma$, $\delta$ to 1.5, 1.0, 1.5, and 0.1.

\section{Experiments}
\subsection{Model Setup}

In our experiments, Qwen2.5-Omni~\cite{Xu2025Qwen25OmniTR} serves as the base model.
After SFT, we perform GRPO training on 4 H100 GPUs. The reward model is served via vLLM. More detailed training settings are provided in the Appendix \ref{sec:traing_setting}.

\paragraph{Data Construction}
We construct our dataset from AVQA~\cite{yang2022avqa} by generating captions with Gemini-2.5-Pro and synthesizing four-dimensional reasoning QA pairs via Qwen3-32B\footnote{\href{https://huggingface.co/Qwen/Qwen3-32B}{https://huggingface.co/Qwen/Qwen3-32B}}~\cite{yang2025qwen3}. We filter for consistency and retain only samples with intermediate difficulty, resulting in 46,544 candidates $Data_{RL}$. A balanced subset of 5,000 is selected for SFT preparation. Qwen3-32B generates CoT paths for this subset, yielding 4,600 high-quality instances $Data_{SFT}$ for Stage-1 supervised fine-tuning (SFT), while the remaining 41,544 samples are reserved for Stage-2 RL.
Detailed generation procedures, prompts, and four dimensions' case studies are presented in Appendix \ref{sec:app_C}.

\subsection{Main Results}
We primarily evaluate model performance on the MMAU and MMAR benchmarks, and further conduct reasoning analysis under the CAFE framework, with benchmark details provided in Appendix \ref{sec:app_E}. MMAU-v05.15.25 and 3B-model results are given in Appendix \ref{sec:s_res}.
From the experimental results in Tables~\ref{table:cafe_results}, \ref{table:benchmark_res} and Figure \ref{fig:cafe_length}, we conclude:

\paragraph{MPAR$^2$ Facilitates Effective Perception Refinement}
As demonstrated in Table~\ref{table:cafe_results} and Figure~\ref{fig:cafe_length}, MPAR$^2$ effectively addresses the challenge of perception decay and boosts perceptual capabilities.
Specifically, in Figure~\ref{fig:cafe_length}, MPAR$^2$ successfully mitigates the trend of perception decay, maintaining both perception and reasoning accuracy at consistently high and stable levels. 
In Table~\ref{table:cafe_results}, MPAR$^2$ achieves leading performance among the evaluated models in terms of perception accuracy $\textit{Acc}_{\text{per}}$ and event utilization error $\textit{Err}_{\text{use}}$, highlighting the potential of our method. 
Meanwhile, to gain deeper insight into perceptual behavior, we conduct more detailed analysis in Appendix \ref{sec:attention} and \ref{sec:decay_analysis}.

\definecolor{ured}{RGB}{228,26,28}
\definecolor{ugreen}{RGB}{77,175,74}

\begin{figure}[tb]
\centering

\begin{minipage}[t]{0.49\linewidth}
\centering
\begin{tikzpicture}
\begin{axis}[
    width=1.2\linewidth,
    height=0.96\linewidth,
    ymajorgrids,
    grid style=dashed,
    xmin=0.5, xmax=10.5,
    ymin=0, ymax=1,
    xtick={1,...,10},
    xticklabels={30,50,70,90,110,130,150,170,190,210},
    ytick={0,0.2,0.4,0.6,0.8,1},
    ylabel={Accuracy},
    ylabel style={
        font=\scriptsize,
        at={(axis description cs:-0.14,0.5)},
        anchor=south,
        overlay
    },
    ticklabel style={font=\tiny},
    label style={font=\scriptsize},
    legend style={
        font=\tiny,
        fill opacity=0.5,
        draw=none,
        at={(axis description cs:0.99,0.99)},
        anchor=north east,
        text opacity=1,
    }
]
\addplot[ured, mark=o,mark size=1.6pt, line width=1pt] coordinates {
    (1,0.370833) (2,0.312323) (3,0.330024) (4,0.315927)
    (5,0.340177) (6,0.247573) (7,0.119444)
    (8,0.057143) (9,0.256410) (10,0.000000)
};
\addplot[ugreen, mark=square, mark size=1.6pt,line width=1pt] coordinates {
    (1,0.575000) (2,0.546875) (3,0.518405) (4,0.527273)
    (5,0.454545) (6,0.295455) (7,0.333333)
    (8,0.400000) (9,0.333333) (10,0.000000)
};
\legend{Perception, Reasoning}
\end{axis}
\end{tikzpicture}

\vspace{-0.4em}
{\small \hspace*{1.6em}Qwen2.5-Omni-7B (Base)}
\end{minipage}
\hfill
\begin{minipage}[t]{0.49\linewidth}
\centering
\begin{tikzpicture}
\begin{axis}[
    width=1.2\linewidth,
    height=0.96\linewidth,
    ymajorgrids,
    grid style=dashed,
    xmin=-1, xmax=19,
    ymin=0, ymax=1,
    xtick={0,2,...,18},
    xticklabels={224,272,320,368,416,464,512,560,608,656},
    ytick={0,0.2,0.4,0.6,0.8,1},
    ticklabel style={font=\tiny},
    label style={font=\scriptsize},
    legend style={
        font=\tiny,
        fill opacity=0.5,
        draw=none,
        at={(axis description cs:0.99,0.99)},
        anchor=north east,
        text opacity=1,
    }
]
\addplot[ured, mark=o, mark size=1.6pt, line width=1pt] coordinates {
   (0,0.436111)
   (2,0.679514)
   (4,0.575277)
   (6,0.639520)
   (8,0.595324)
   (10,0.624339)
   (12,0.642732)
   (14,0.644432)
   (16,0.679819)
   (18,0.590235)
};

\addplot[ugreen, mark=square, mark size=1.6pt, line width=1pt] coordinates {
   (0,0.833333)
   (2,0.833333)
   (4,0.581395)
   (6,0.476744)
   (8,0.542056)
   (10,0.542636)
   (12,0.540000)
   (14,0.603175)
   (16,0.612245)
   (18,0.640625)
};
\legend{Perception, Reasoning}
\end{axis}
\end{tikzpicture}

\vspace{-0.4em}
{\small \hspace*{1.6em}MPAR$^2$-7B}
\end{minipage}

\caption{
Comparison of accuracy across token lengths.}
\label{fig:cafe_length}
\vspace{-10pt}
\end{figure}

\paragraph{MPAR$^2$ Stably Improves Perception and Reasoning Capabilities}
As shown in Table~\ref{table:benchmark_res}, compared to the Qwen2.5-Omni-7B baseline, MPAR$^2$ achieves substantial performance gains. 
Specifically, MPAR$^2$ boosts accuracy on MMAU from 65.90\% to 74.59\% and on MMAR from 55.20\% to 60.32\%. 
Notably, stable improvements are observed across distinct audio domains, which supports our core insight: \textit{enhanced perception during reasoning can suppress audio perception decay and boost reasoning performance.}
However, the gain on MMAR is smaller than on MMAU, and MPAR$^2$ remains behind Step-Audio-R1, while achieving higher perception accuracy on CAFE.
We attribute these differences to two factors that jointly affect the reasoning performance of LALMs: \textit{audio perception capacity during reasoning} and \textit{inherent logical reasoning capacity}.
MPAR$^2$'s main goal is to strengthen perceptual anchoring throughout the entire reasoning process, with greater emphasis on the former factor.
For stronger logical ability, we will further explore data and ways to enhance text-based reasoning ability in future work.

\section{Analysis}
\paragraph{Ablation Study} 
We show the impact of different reward ablations in Table~\ref{table:benchmark_res}, demonstrating that each mechanism contributes positively to MPAR$^2$. 
First, relying solely on format and simple accuracy rewards already yields substantial gains of 6.91\% on MMAU and 2.32\% on MMAR over the baseline, validating the effectiveness of the MPAR$^2$ pattern. 
Second, incorporating the review-enhanced accuracy mechanism leads to further improvements across all metrics, underscoring the robustness and efficacy of the rethinking process. 
Finally, the integration of perception and multi-step reasoning rewards specifically boosts perception capabilities, elevating MMAU performance from 73.36\% and 74.08\% to 74.59\%. 
Notably, the perception reward contributes a relatively larger gain, highlighting the critical role of perception in audio reasoning.
Meanwhile, More ablations on model size and method are provided in Appendix \ref{sec_c_3b} and \ref{sec_ablation}.

\usetikzlibrary{decorations.markings}
\definecolor{ugreen}{RGB}{098,190,166}
\definecolor{uyellow}{RGB}{253,186,107}
\definecolor{ured}{RGB}{235,096,070}
\definecolor{upurple}{RGB}{175,135,220} 
\definecolor{ublue}{RGB}{076,135,220}
\definecolor{ugray}{RGB}{150,156,165}
\definecolor{ubrown}{RGB}{176,125,086}

\pgfmathsetlengthmacro{\BarW}{12pt}
\pgfmathsetlengthmacro{\BarStep}{16pt}
\pgfmathsetlengthmacro{\GapClusters}{-10pt}

\pgfmathsetlengthmacro{\ShiftEasy}{-2.0*\BarStep - 0.5*\GapClusters}
\pgfmathsetlengthmacro{\ShiftMedium}{ -1.0*\BarStep - 0.5*\GapClusters}
\pgfmathsetlengthmacro{\ShiftSynth}{-2.5*\BarStep}
\pgfmathsetlengthmacro{\ShiftHard}{0.0*\BarStep - 0.5*\GapClusters}
\pgfmathsetlengthmacro{\ShiftIE}{1.0*\BarStep - 0.5*\GapClusters}
\pgfmathsetlengthmacro{\ShiftReason}{2.0*\BarStep - 0.5*\GapClusters}

\pgfmathsetlengthmacro{\ShiftSingle}{ -1.5*\BarStep + 0.5*\GapClusters}
\pgfmathsetlengthmacro{\ShiftMixture}{-0.5*\BarStep + 0.5*\GapClusters}
\pgfmathsetlengthmacro{\ShiftPercep}{  0.5*\BarStep + 0.5*\GapClusters}
\pgfmathsetlengthmacro{\ShiftHigh}{    1.5*\BarStep + 0.5*\GapClusters}

\tikzset{
  hatchbar/.style={
    postaction={draw=none, pattern=north east lines}
  },
  hatchCross/.style={
    postaction={draw=none, pattern=crosshatch}
  }
}

\begin{figure}[t]
\centering
\resizebox{\linewidth}{!}{%
\begin{tikzpicture}

\begin{axis}[
  ybar,
  bar width=\BarW,
  width=1.08\linewidth,
  height=0.66\linewidth,
  ymin=360, ymax=680,
  ytick={400, 440, 480, 520, 560, 600, 640},
  symbolic x coords={MMAU (Test-mini),MMAR},
  xtick={MMAU (Test-mini),MMAR},
  enlarge x limits=0.52,
  axis line style={thick},
  tick style={thick,draw=none},
  label style={font=\normalsize},
  yticklabel style={font=\scriptsize},
  xticklabel style={font=\footnotesize},
  ymajorgrids,
  grid style={dashed, gray!30},
  legend style={
    font=\small,
    draw=none,
    fill=none,
    text opacity=1,
    rounded corners,
    at={(-0.05,1.38)},
    anchor=north west,
    legend cell align=left,
    nodes={anchor=mid west},
    /tikz/every node/.append style={yshift=2pt}
  },
  legend image code/.code={
    \path[#1, draw=none, fill opacity=0.95] (0pt,-1pt) rectangle (8pt,7pt);
  },
  legend columns=3,
  transpose legend, 
  clip=false,
]

\addplot+[forget plot, draw=uyellow!80, fill=uyellow!40, bar shift=\ShiftEasy]
  coordinates {(MMAU (Test-mini),388.90)};

\addplot+[forget plot, draw=uyellow!80, fill=uyellow!70, bar shift=\ShiftMedium]
  coordinates {(MMAU (Test-mini),453.99)};

\addplot+[forget plot, draw=uyellow!80, fill=uyellow!90, bar shift=\ShiftHard]
  coordinates {(MMAU (Test-mini),473.34)};

\addplot+[forget plot, draw=ublue!80, fill=ublue!50, bar shift=\ShiftIE]
  coordinates {(MMAU (Test-mini),402.98)};

\addplot+[forget plot, draw=ublue!80, fill=ublue!80, bar shift=\ShiftReason]
  coordinates {(MMAU (Test-mini),463.54)};

\addplot+[forget plot, draw=ugreen!60, fill=ugreen!30, bar shift=\ShiftSingle]
  coordinates {(MMAR,578.186)};

\addplot+[forget plot, draw=ugreen!60, fill=ugreen!50, bar shift=\ShiftMixture]
  coordinates {(MMAR,633.5275)};

\addplot+[forget plot, draw=ured!80, fill=ured!50, bar shift=\ShiftPercep]
  coordinates {(MMAR,561.28)};

\addplot+[forget plot, draw=ured!80, fill=ured!70, bar shift=\ShiftHigh]
  coordinates {(MMAR,614.37)};

\addlegendimage{ybar, draw=uyellow!80, fill=uyellow!40}
\addlegendentry{Easy}

\addlegendimage{ybar, draw=uyellow!80, fill=uyellow!70}
\addlegendentry{Medium}

\addlegendimage{ybar, draw=uyellow!80, fill=uyellow!90}
\addlegendentry{Hard}

\addlegendimage{ybar, draw=ublue!80, fill=ublue!50}
\addlegendentry{Information Extraction}

\addlegendimage{ybar, draw=ublue!80, fill=ublue!80}
\addlegendentry{Reasoning}

\addlegendimage{ybar, draw=ugreen!80, fill=ugreen!30}
\addlegendentry{Single-Source}

\addlegendimage{ybar, draw=ugreen!80, fill=ugreen!50 }
\addlegendentry{Acoustic Mixtures}

\addlegendimage{ybar, draw=ured!80, fill=ured!50}
\addlegendentry{Perception}

\addlegendimage{ybar, draw=ured!80, fill=ured!70}
\addlegendentry{High-level Cognitive}

\node[font=\normalsize, text=red!60!black]
  at ([xshift=-86pt]axis cs:MMAR,600)
  {\textbf{Reasoning Length Gap}};
  
\draw[<->, very thick]
    ([xshift=\ShiftSynth+11pt]axis cs:MMAR,578.186) --
    ([xshift=\ShiftSynth+11pt]axis cs:MMAR,633.5275)
    node[midway, above=8pt, font=\scriptsize] {$\boldsymbol{\Delta=55}$};
    

\draw[<->, very thick]
    ([xshift=\ShiftSynth+43pt]axis cs:MMAR,561.28) --
    ([xshift=\ShiftSynth+43pt]axis cs:MMAR,614.37)
    node[midway, above=8pt, font=\scriptsize, xshift=0.3cm] {$\boldsymbol{\Delta=53}$};

\draw[<->, very thick]
    ([xshift=\ShiftSynth-81pt]axis cs:MMAR,388.90) --
    ([xshift=\ShiftSynth-81pt]axis cs:MMAR,453.99)
    node[midway, above=10pt, font=\scriptsize] {$\boldsymbol{\Delta=65}$};

\draw[<->, very thick]
    ([xshift=\ShiftSynth-33pt]axis cs:MMAR,402.98) --
    ([xshift=\ShiftSynth-33pt]axis cs:MMAR,463.54)
    node[midway, above=11pt, font=\scriptsize] {$\boldsymbol{\Delta=61}$};
    
\draw[thick]
  (rel axis cs:0,0) -- (rel axis cs:1,0);

\end{axis}
\end{tikzpicture}
}

\caption{Token length of MPAR$^2$-7B by task category on MMAU and MMAR. The y-axis reports tokens.}
\label{fig:length}
\vspace{-10pt}
\end{figure}

\paragraph{Adaptive Reasoning Budget and Latency Trade-off}
As illustrated in Figure \ref{fig:length}, we summarized the reasoning token lengths of MPAR$^2$ across different aspects on benchmarks. 
Notably, since MMAR focuses on deep reasoning problems, the model exhibits a longer overall output length compared to MMAU. 
Beyond this dataset-level observation, we find that straightforward inputs, such as easy-level questions or basic information extraction tasks, lead to shorter reasoning, whereas more complex problems result in longer responses.
This behavior suggests that our MPAR$^2$ successfully enables the model to dynamically adjust its reasoning budget rather than following a fixed long CoT pattern. 
Specifically, it reduces the reasoning budget for simple tasks while extending it for complex scenarios, the model reduces the unnecessary overhead introduced by CoT in simple audio QA tasks, achieving a better alignment between reasoning budget and task complexity.

\section{Conclusion}
In this work, we first investigate why RL for structured reasoning yields marginal or even negative gains over RL for direct answering.
Based on the results obtained from the proposed CAFE framework, we identify the critical bottleneck: as reasoning scales, reasoning capability suffers from audio perception decay.
To address it, we propose MPAR$^2$, which successfully mitigates perception decay, improves perceptual performance during reasoning, further enhancing overall reasoning ability.
Additionally, in-depth analysis shows that MPAR$^2$ optimizes audio attention during reasoning and dynamically adjusts reasoning budget allocation.

\section*{Limitations}
\paragraph{Model Size Limit}
Due to GPU constraints, our experiments are limited to models up to 7B. Scaling to larger models is left to our future work.
\paragraph{Depth of Logical Reasoning}
Our dataset and method focus on perceptual enhancement during reasoning, not on deeper logical reasoning itself. Extending to harder QA data and methods that target inherent reasoning ability is our future work.
\paragraph{Reasoning Depth vs. Real-Time Latency}
In spoken question answering, test-time scaling improves reasoning quality at the cost of higher latency, creating a fundamental tension for dialogue systems that demand fast and responsive interaction. Traditional sequential pipelines follow a listen, think, speak paradigm that preserves accuracy but incurs substantial delay.
Future work will explore more efficient reasoning and dynamic scheduling of reasoning and response generation to better balance accuracy and responsiveness.

\section*{Acknowledgements}
This work was supported in part by the National Science Foundation of China (Nos. 62276056 and U24A20334), the Yunnan Fundamental Research Projects (No.202401BC070021), the Yunnan Science and Technology Major Project (No. 202502AD080014), the Liaoning Provincial Science and Technology Plan Project (No. 2026JH40/10100033), the Fundamental Research Funds for the Central Universities (Nos. N25BSS054 and N25BSS094), and the Program of Introducing Talents of Discipline to Universities, Plan 111 (No.B16009).

\vspace{-10pt}
\bibliography{references}

@article{snell2024scaling,
  title={Scaling llm test-time compute optimally can be more effective than scaling model parameters},
  author={Snell, Charlie and Lee, Jaehoon and Xu, Kelvin and Kumar, Aviral},
  journal={arXiv preprint arXiv:2408.03314},
  year={2024}
}

@article{guo2025deepseek,
  title={Deepseek-r1: Incentivizing reasoning capability in llms via reinforcement learning},
  author={Guo, Daya and Yang, Dejian and Zhang, Haowei and Song, Junxiao and Zhang, Ruoyu and Xu, Runxin and Zhu, Qihao and Ma, Shirong and Wang, Peiyi and Bi, Xiao and others},
  journal={arXiv preprint arXiv:2501.12948},
  year={2025}
}

@article{ma2025mmar,
  title={MMAR: A Challenging Benchmark for Deep Reasoning in Speech, Audio, Music, and Their Mix},
  author={Ma, Ziyang and Ma, Yinghao and Zhu, Yanqiao and Yang, Chen and Chao, Yi-Wen and Xu, Ruiyang and Chen, Wenxi and Chen, Yuanzhe and Chen, Zhuo and Cong, Jian and others},
  journal={arXiv preprint arXiv:2505.13032},
  year={2025}
}

@article{sakshi2024mmau,
  title={Mmau: A massive multi-task audio understanding and reasoning benchmark},
  author={Sakshi, S and Tyagi, Utkarsh and Kumar, Sonal and Seth, Ashish and Selvakumar, Ramaneswaran and Nieto, Oriol and Duraiswami, Ramani and Ghosh, Sreyan and Manocha, Dinesh},
  journal={arXiv preprint arXiv:2410.19168},
  year={2024}
}

@article{tian2025step,
  title={Step-Audio-R1 Technical Report},
  author={Tian, Fei and Zhang, Xiangyu Tony and Zhang, Yuxin and Zhang, Haoyang and Li, Yuxin and Liu, Daijiao and Deng, Yayue and Wu, Donghang and Chen, Jun and Zhao, Liang and others},
  journal={arXiv preprint arXiv:2511.15848},
  year={2025}
}

@article{li2025reinforcement,
  title={Reinforcement learning outperforms supervised fine-tuning: A case study on audio question answering},
  author={Li, Gang and Liu, Jizhong and Dinkel, Heinrich and Niu, Yadong and Zhang, Junbo and Luan, Jian},
  journal={arXiv preprint arXiv:2503.11197},
  year={2025}
}

@article{wu2025audio,
  title={Audio-thinker: Guiding audio language model when and how to think via reinforcement learning},
  author={Wu, Shu and Li, Chenxing and Wang, Wenfu and Zhang, Hao and Wang, Hualei and Yu, Meng and Yu, Dong},
  journal={arXiv preprint arXiv:2508.08039},
  year={2025}
}

@article{rouditchenko2025omni,
  title={Omni-R1: Do You Really Need Audio to Fine-Tune Your Audio LLM?},
  author={Rouditchenko, Andrew and Bhati, Saurabhchand and Araujo, Edson and Thomas, Samuel and Kuehne, Hilde and Feris, Rogerio and Glass, James},
  journal={arXiv preprint arXiv:2505.09439},
  year={2025}
}

@article{Xu2025Qwen25OmniTR,
  title={Qwen2.5-Omni Technical Report},
  author={Jin Xu and Zhifang Guo and Jinzheng He and Hangrui Hu and Ting He and Shuai Bai and Ke-qin Chen and Jialin Wang and Yang Fan and Kai Dang and Bin Zhang and Xiong Wang and Yunfei Chu and Junyang Lin},
  journal={ArXiv},
  year={2025},
  volume={abs/2503.20215},
  url={https://api.semanticscholar.org/CorpusID:277322543}
}

@article{chu2024qwen2,
  title={Qwen2-audio technical report},
  author={Chu, Yunfei and Xu, Jin and Yang, Qian and Wei, Haojie and Wei, Xipin and Guo, Zhifang and Leng, Yichong and Lv, Yuanjun and He, Jinzheng and Lin, Junyang and others},
  journal={arXiv preprint arXiv:2407.10759},
  year={2024}
}

@article{zheng2023judging,
  title={Judging llm-as-a-judge with mt-bench and chatbot arena},
  author={Zheng, Lianmin and Chiang, Wei-Lin and Sheng, Ying and Zhuang, Siyuan and Wu, Zhanghao and Zhuang, Yonghao and Lin, Zi and Li, Zhuohan and Li, Dacheng and Xing, Eric and others},
  journal={Advances in neural information processing systems},
  volume={36},
  pages={46595--46623},
  year={2023}
}

@article{zhang2025mimo,
  title={MiMo-Audio: Audio Language Models are Few-Shot Learners},
  author={Zhang, Dong and Wang, Gang and Xue, Jinlong and Fang, Kai and Zhao, Liang and Ma, Rui and Ren, Shuhuai and Liu, Shuo and Guo, Tao and Zhuang, Weiji and others},
  journal={arXiv preprint arXiv:2512.23808},
  year={2025}
}

@article{lu2026desta2,
  title={Desta2. 5-audio: Toward general-purpose large audio language model with self-generated cross-modal alignment},
  author={Lu, Ke-Han and Chen, Zhehuai and Fu, Szu-Wei and Yang, Chao-Han Huck and Huang, Sung-Feng and Yang, Chih-Kai and Yu, Chee-En and Chen, Chun-Wei and Chen, Wei-Chih and Huang, Chien-yu and others},
  journal={IEEE Transactions on Audio, Speech and Language Processing},
  year={2026},
  publisher={IEEE}
}

@article{abouelenin2025phi,
  title={Phi-4-mini technical report: Compact yet powerful multimodal language models via mixture-of-loras},
  author={Abouelenin, Abdelrahman and Ashfaq, Atabak and Atkinson, Adam and Awadalla, Hany and Bach, Nguyen and Bao, Jianmin and Benhaim, Alon and Cai, Martin and Chaudhary, Vishrav and Chen, Congcong and others},
  journal={arXiv preprint arXiv:2503.01743},
  year={2025}
}

@article{goel2025audio,
  title={Audio flamingo 3: Advancing audio intelligence with fully open large audio language models},
  author={Goel, Arushi and Ghosh, Sreyan and Kim, Jaehyeon and Kumar, Sonal and Kong, Zhifeng and Lee, Sang-gil and Yang, Chao-Han Huck and Duraiswami, Ramani and Manocha, Dinesh and Valle, Rafael and others},
  journal={arXiv preprint arXiv:2507.08128},
  year={2025}
}

@article{xie2025audio,
  title={Audio-reasoner: Improving reasoning capability in large audio language models},
  author={Xie, Zhifei and Lin, Mingbao and Liu, Zihang and Wu, Pengcheng and Yan, Shuicheng and Miao, Chunyan},
  journal={arXiv preprint arXiv:2503.02318},
  year={2025}
}

@article{comanici2025gemini,
  title={Gemini 2.5: Pushing the frontier with advanced reasoning, multimodality, long context, and next generation agentic capabilities},
  author={Comanici, Gheorghe and Bieber, Eric and Schaekermann, Mike and Pasupat, Ice and Sachdeva, Noveen and Dhillon, Inderjit and Blistein, Marcel and Ram, Ori and Zhang, Dan and Rosen, Evan and others},
  journal={arXiv preprint arXiv:2507.06261},
  year={2025}
}

@article{hurst2024gpt,
  title={Gpt-4o system card},
  author={Hurst, Aaron and Lerer, Adam and Goucher, Adam P and Perelman, Adam and Ramesh, Aditya and Clark, Aidan and Ostrow, AJ and Welihinda, Akila and Hayes, Alan and Radford, Alec and others},
  journal={arXiv preprint arXiv:2410.21276},
  year={2024}
}

@inproceedings{yang2022avqa,
  title={Avqa: A dataset for audio-visual question answering on videos},
  author={Yang, Pinci and Wang, Xin and Duan, Xuguang and Chen, Hong and Hou, Runze and Jin, Cong and Zhu, Wenwu},
  booktitle={Proceedings of the 30th ACM international conference on multimedia},
  pages={3480--3491},
  year={2022}
}

@article{yang2025qwen3,
  title={Qwen3 technical report},
  author={Yang, An and Li, Anfeng and Yang, Baosong and Zhang, Beichen and Hui, Binyuan and Zheng, Bo and Yu, Bowen and Gao, Chang and Huang, Chengen and Lv, Chenxu and others},
  journal={arXiv preprint arXiv:2505.09388},
  year={2025}
}

@article{shao2024deepseekmath,
  title={Deepseekmath: Pushing the limits of mathematical reasoning in open language models},
  author={Shao, Zhihong and Wang, Peiyi and Zhu, Qihao and Xu, Runxin and Song, Junxiao and Bi, Xiao and Zhang, Haowei and Zhang, Mingchuan and Li, YK and Wu, Y and others},
  journal={arXiv preprint arXiv:2402.03300},
  year={2024}
}

@article{marjanovic2025deepseek,
  title={DeepSeek-R1 Thoughtology: Let's think about LLM Reasoning},
  author={Marjanovi{\'c}, Sara Vera and Patel, Arkil and Adlakha, Vaibhav and Aghajohari, Milad and BehnamGhader, Parishad and Bhatia, Mehar and Khandelwal, Aditi and Kraft, Austin and Krojer, Benno and L{\`u}, Xing Han and others},
  journal={arXiv preprint arXiv:2504.07128},
  year={2025}
}

@article{lee2025evaluating,
  title={Evaluating step-by-step reasoning traces: A survey},
  author={Lee, Jinu and Hockenmaier, Julia},
  journal={arXiv preprint arXiv:2502.12289},
  year={2025}
}

@article{mcalister1879xiii,
  title={XIII. The law of the geometric mean},
  author={McAlister, Donald},
  journal={Proceedings of the Royal Society of London},
  volume={29},
  number={196-199},
  pages={367--376},
  year={1879},
  publisher={The Royal Society London}
}

@article{ma2025audio,
  title={Audio-cot: Exploring chain-of-thought reasoning in large audio language model},
  author={Ma, Ziyang and Chen, Zhuo and Wang, Yuping and Chng, Eng Siong and Chen, Xie},
  journal={arXiv preprint arXiv:2501.07246},
  year={2025}
}

@article{wen2025sari,
  title={Sari: Structured audio reasoning via curriculum-guided reinforcement learning},
  author={Wen, Cheng and Guo, Tingwei and Zhao, Shuaijiang and Zou, Wei and Li, Xiangang},
  journal={arXiv preprint arXiv:2504.15900},
  year={2025}
}

@incollection{benesty2009pearson,
  title={Pearson correlation coefficient},
  author={Benesty, Jacob and Chen, Jingdong and Huang, Yiteng and Cohen, Israel},
  booktitle={Noise reduction in speech processing},
  pages={1--4},
  year={2009},
  publisher={Springer}
}

@article{chang2024efficient,
  title={Efficient prompting methods for large language models: A survey},
  author={Chang, Kaiyan and Xu, Songcheng and Wang, Chenglong and Luo, Yingfeng and Liu, Xiaoqian and Xiao, Tong and Zhu, Jingbo},
  journal={arXiv preprint arXiv:2404.01077},
  year={2024}
}

@article{wei2022chain,
  title={Chain-of-thought prompting elicits reasoning in large language models},
  author={Wei, Jason and Wang, Xuezhi and Schuurmans, Dale and Bosma, Maarten and Xia, Fei and Chi, Ed and Le, Quoc V and Zhou, Denny and others},
  journal={Advances in neural information processing systems},
  volume={35},
  pages={24824--24837},
  year={2022}
}

@article{brown2024large,
  title={Large language monkeys: Scaling inference compute with repeated sampling},
  author={Brown, Bradley and Juravsky, Jordan and Ehrlich, Ryan and Clark, Ronald and Le, Quoc V and R{\'e}, Christopher and Mirhoseini, Azalia},
  journal={arXiv preprint arXiv:2407.21787},
  year={2024}
}

@article{jaech2024openai,
  title={Openai o1 system card},
  author={Jaech, Aaron and Kalai, Adam and Lerer, Adam and Richardson, Adam and El-Kishky, Ahmed and Low, Aiden and Helyar, Alec and Madry, Aleksander and Beutel, Alex and Carney, Alex and others},
  journal={arXiv preprint arXiv:2412.16720},
  year={2024}
}

@article{lee2025probing,
  title={Probing the Difficulty Perception Mechanism of Large Language Models},
  author={Lee, Sunbowen and Yin, Qingyu and Leong, Chak Tou and Zhang, Jialiang and Gong, Yicheng and Ni, Shiwen and Yang, Min and Shen, Xiaoyu},
  journal={arXiv preprint arXiv:2510.05969},
  year={2025}
}

@article{dai2025s,
  title={S-GRPO: Early Exit via Reinforcement Learning in Reasoning Models},
  author={Dai, Muzhi and Yang, Chenxu and Si, Qingyi},
  journal={arXiv preprint arXiv:2505.07686},
  year={2025}
}

@article{waheed2025less,
  title={Less is More Tokens: Efficient Math Reasoning via Difficulty-Aware Chain-of-Thought Distillation},
  author={Waheed, Abdul and Mitra, Chancharik and Wang, Laurie Z and Ramanan, Deva and Raj, Bhiksha},
  journal={arXiv preprint arXiv:2509.05226},
  year={2025}
}

@article{fan2025incentivizing,
  title={Incentivizing Consistent, Effective and Scalable Reasoning Capability in Audio LLMs via Reasoning Process Rewards},
  author={Fan, Jiajun and Ren, Roger and Li, Jingyuan and Pandey, Rahul and Shivakumar, Prashanth Gurunath and Bulyko, Ivan and Gandhe, Ankur and Liu, Ge and Gu, Yile},
  journal={arXiv preprint arXiv:2510.20867},
  year={2025}
}
\clearpage
\section*{Appendix}
\appendix
\definecolor{PromptBoxTitleColor}{RGB}{76,76,76}
\definecolor{PromptBoxColor}{RGB}{247,247,255}
\definecolor{block-gray}{gray}{0.85}

\newtcolorbox{mybackground}{
  colback=block-gray!35,
  grow to right by=0mm,
  grow to left by=0mm,
  boxrule=0pt,
  arc=1pt,
  left=3pt,
  right=3pt,
  top=3pt,
  bottom=3pt,
  breakable,
  fontupper=\small
}

\section{CAFE Framework Prompts}
\label{sec:app_A}
In terms of prompt design, we have referred to efficient prompting methods~\cite{chang2024efficient}.
The following are the Gemini-3-Pro caption and audio event extraction prompts used by the CAFE framework.

\subsection{Gemini-3-Pro Caption prompt}
\begin{mybackground}
Please generate a detailed chronological description of the following audio clip.\\

Listen carefully to the sequence of events and describe the audio flow from beginning to end. Your caption should:

- Explicitly state the order of sounds using transitional phrases (e.g., The audio begins with..., Followed by..., Simultaneously..., As the sound fades...).

- Capture the subtle details of each sound event, including its duration and intensity changes.

- Distinguish between foreground events (main actions) and background noise (ambience).\\

Output the result as a cohesive narrative text without line breaks or bullet points.
\end{mybackground}

\subsection{Audio Event Extraction prompt}
\begin{mybackground}
\textbf{Role}

You are an expert Audio Logic Consistency Evaluator. Your task is to evaluate a Model Reasoning Path against the Question, Correct Answer, and Ground Truth Caption.\\

\textbf{Step 1: Analyze Requirements}

1. Determine `required\_events`: The essential sounds from the [Ground Truth Audio Caption] needed to answer the [Question] correctly.

2. Identify `all\_caption\_events`: All sounds actually present in the audio.\\

\textbf{Step 2: Categorize Model Events}

For every audio event mentioned in the [Model Reasoning Path], categorize it into ONE of the following 4 lists based on its Validity (Is it real?) and Usage (How did the model use it?):

\textbf{1. matched\_events} (Correct \& Necessary):

 - The event is in `required\_events`.
 
 - The model used it effectively to derive the answer.

\textbf{2. error\_matched} (Fabrication / Misidentification):

 - The event is NOT in `all\_caption\_events` (Hallucination).
 
 - OR The model heard sound A (real) but identified it as sound B (fake/wrong), and B is not in the caption.
 
 - These are False Positives regarding perception.

\textbf{3. error\_use} (Distraction / Wrong Reasoning):

 - The event exists in `all\_caption\_events` but is NOT in `required\_events` (Irrelevant).
 
 - CRITICAL CONDITION: The model activley used this irrelevant sound to support a conclusion, or the model was confused by it.
 
 - Example: I hear a bird (irrelevant), so the answer must be 'Forest'. (When the answer is actually 'Park' due to other sounds).

\textbf{4. neutral\_events} (Harmless Mention / Valid Filtering):

 - The event exists in 'all\_caption\_events' but is NOT in `required\_events`.
 
 - The model mentioned it only to describe the scene or explicitly stated it was not relevant.
 
 - Example: I hear wind in the background, but the main sound is the car engine. (Here, 'wind' is a neutral mention, not an error).
 
 - Action: Do NOT count these as errors.

\textbf{5. missed\_events} (Omission):

 - Events in `required\_events` that are NOT found in `matched\_events`.\\

\textbf{Input Data}

[Question]:
{{QUESTION}}

[Correct Answer]:
{{CORRECT\_ANSWER}}

[Ground Truth Audio Caption]:\\
{{GROUND\_TRUTH\_CAPTION}}

[Model Reasoning Path]:
{{MODEL\_REASONING}}\\

\textbf{Output Format} (JSON Only)
{

all\_reasoning\_events: [list of all events mentioned by model],

matched\_events: [list],

error\_matched: [list],

error\_use: [list],

neutral\_events: [list],

missed\_events: [list]

}
\end{mybackground}

\section{CAFE Evaluation Supplement}
\label{sec:app_B}

\subsection{Test model}
In Section~\ref{sec:cafe}, the open-source LALMs evaluated include MiMo-Audio~\citep{zhang2025mimo}, Qwen2.5-Omni-7B~\citep{Xu2025Qwen25OmniTR}, DeSTA2.5-Audio~\citep{lu2026desta2}, Phi-4-Multimodal~\citep{abouelenin2025phi}, Qwen2-Audio-Instruct~\citep{chu2024qwen2}, as well as Omni-R1~\citep{rouditchenko2025omni} without Chain-of-Thought (CoT) RL training. 
Regarding LARMs, we evaluate Audio-Flamingo-3~\citep{goel2025audio}, Audio Reasoner~\citep{xie2025audio}, Step-Audio-R1.1 and Step-Audio-R1~\citep{tian2025step}. Additionally, we incorporate the commercial models Gemini-2.5-Flash~\citep{comanici2025gemini} and GPT-4o-Audio~\citep{hurst2024gpt}. 

\subsection{Model Inference Prompt}
In CAFE evaluation experiments (Figure \ref{fig:six_plots}), models (c) and (d) lack CoT training, their output lengths and formats are inherently unstable. We use the "answer first, then reasoning" prompt strictly to ensure stable answer extraction. To eliminate any concerns and guarantee an absolutely fair comparison, we re-evaluated the baseline LALMs using a "reasoning first, then answer" prompt. The results demonstrate that the models still exhibit clear Audio Perception Decay. Furthermore, the overall reasoning accuracy of the baselines did not change meaningfully.

The following prompt is used to evaluate LALMs and commercial models:
\begin{mybackground}
[Question]

Output the selected option first. Then write Reasoning: and explain your choice, mentioning the audio events that support it.

[Choices]
\end{mybackground}

Prompt used to evaluate LARMs:
\begin{mybackground}
[Question]

Select one option from the provided choices.

[Choices]
\end{mybackground}

\subsection{CAFE Framework Reliability Verification}
\label{sec:CAFE_Verification}
Following the LLM-as-a-Judge paradigm, we employ Gemini-3-Pro for CAFE’s scalable evaluation, leveraging its strong multimodal capabilities. To further assess the reliability of the automatic evaluation, we conduct human cross-validation on 100 randomly sampled instances. Specifically, human annotators verify both the caption text and the events extracted by Gemini-3-Pro, where the extracted events are evaluated across the categories of Error-matched, Error-used, Matched, Missed, and Neutral. The category-wise human--Gemini agreement is reported in Table~\ref{table:CAFE_ver} , with an average agreement rate of 92\%, indicating that Gemini-3-Pro provides sufficiently reliable judgments for our evaluation setting.

To assess cross-judge and prompt stability, we replace Gemini-3-Pro with GPT-5 as the CAFE judge and repeat the evaluation using semantically equivalent prompts. Perception accuracy changes by less than 2.1 percentage points, while the relative rankings of all evaluated models remain unchanged. Together with the 92\% human–Gemini agreement, these results indicate that the main CAFE trends are stable and are not an artifact of a single judge model or prompt wording.

\subsection{Model Experiment Detail}
\label{sec:cafe_detail}
Our experiment classifies model responses post-hoc based on their generated length. In Figure \ref{fig:six_plots}, models (a) and (b) underwent CoT training to enhance their performance on high-difficulty questions. Models (c) and (d) did not receive CoT training. However, exactly like models (a) and (b), they exhibit a clear decay in audio perception accuracy as reasoning length increases. We attribute this uniformly to an insufficient perception capacity when processing complex tasks through step-by-step reasoning. This deficiency directly damages reasoning accuracy as the reasoning chain elongates.

Furthermore, we conducted a comparative experiment on the non-CoT trained models (c) and (d) without enforcing any reasoning output constraints. We found that their final accuracy remained fundamentally consistent with the CoT reasoning results, showing only a minor fluctuation of approximately ±0.8\%.

\begin{table}[t]
    \centering
    \renewcommand{\arraystretch}{1}
    \setlength{\tabcolsep}{2pt}
    \begin{tabular}{rrrrrrrr}
        \toprule
         & Cap. & E-mat. & E-use. & Mat. & Miss. & Neu. & Avg. \\
        \midrule
        Agr. & 94\% & 89\% & 91\% & 92\% & 93\% & 93\% & 92\% \\
        \bottomrule
    \end{tabular}
    \caption{Human agreement rates on 100 randomly sampled instances. The table reports agreement (Agr.) for the caption text (Cap.) and for events extracted by Gemini-3-Pro across the categories Error-matched (E-mat.), Error-used (E-use.), Matched (Mat.), Missed (Miss.), and Neutral (Neu.). The last column shows the mean agreement across all reported items.}
    \label{table:CAFE_ver}
\end{table}


\subsection{Difficulty-Stratified Analysis}
Our claim is not that long reasoning is inherently harmful, but that extended reasoning without sufficient audio grounding is associated with, and may amplify, perception errors. In audio reasoning, question difficulty and response length are naturally correlated: harder questions generally require more perceptual checks and reasoning steps, and therefore tend to produce longer responses. To examine whether the decay is merely a by-product of this difficulty effect, we further decouple the two factors through a difficulty-stratified analysis.

Specifically, we split each model's responses into Short, Medium, and Long buckets within each difficulty group, following the length-bucketing setup used in Figure \ref{fig:six_plots}. Hard examples exhibit limited within-group length separation and are therefore omitted from this analysis. As shown in Table~\ref{table:difficulty_length_cafe}, perception accuracy consistently decreases as response length increases within both the Easy and Medium groups across all six evaluated models. This result indicates that the observed decay is not merely caused by harder questions producing longer answers. Even among questions of comparable difficulty, models are more likely to lose stable acoustic anchoring as their generated reasoning becomes longer.

\begin{table}[t]
    \centering
    \small
    \renewcommand{\arraystretch}{1}
    \setlength{\tabcolsep}{2pt}
    \begin{tabular}{lrrrrrr}
        \toprule
        \textbf{Diff.--Len.}
        & \textbf{AR}
        & \textbf{DeSTA}
        & \textbf{MiMo}
        & \textbf{Q2A}
        & \textbf{Q2.5O}
        & \textbf{SAR1} \\
        \midrule
        Easy--S   & 34.39 & 32.57 & 58.13 & 17.01 & 41.56 & 61.07 \\
        Easy--M   & 31.24 & 24.98 & 43.94 & 11.77 & 36.20 & 56.46 \\
        Easy--L   & 26.65 & 20.76 & 38.19 &  6.54 & 31.72 & 51.13 \\
        \midrule
        Medium--S & 34.57 & 29.68 & 55.83 & 14.68 & 37.29 & 60.65 \\
        Medium--M & 30.21 & 24.11 & 45.43 &  8.01 & 26.50 & 53.51 \\
        Medium--L & 23.11 & 21.21 & 37.91 &  6.06 & 21.95 & 48.27 \\
        \bottomrule
    \end{tabular}
    \caption{Difficulty-stratified CAFE perception accuracy (\%) across response-length buckets. S, M, and L denote Short, Medium, and Long responses, respectively. AR: Audio-Reasoner; DeSTA: DeSTA2.5-Audio; MiMo: MiMo-Audio; Q2A: Qwen2-Audio; Q2.5O: Qwen2.5-Omni; SAR1: Step-Audio-R1. Hard examples are omitted because they exhibit limited within-group length separation.}
    \label{table:difficulty_length_cafe}
\end{table}

\subsection{Reasoning Prompts and Response Cases}
\label{sec:prompts_rl}
Implicit and explicit unstructured reasoning prompts for RL:
\\

\begin{mybackground}
\textbf{Implicit Prompt}: \\
Output the final answer in \textless ANSWER \textgreater \textless /ANSWER\textgreater \\
\\
\textbf{Explicit Unstructured Reasoning Prompt}: \\
Output the thinking process in \textless THINK\textgreater \textless/THINK\textgreater and final answer in \textless ANSWER\textgreater \textless /ANSWER\textgreater.
\end{mybackground}

Response cases from different models:
\begin{mybackground}
\textbf{Implicit Case:} \\
Instruction: \\
Based on the given audio, identify the source of the whoop. Please choose the answer from
the following options: [’Human’, ’Bird’, ’Dog’, ’Machine’].\\
Output the final answer in <ANSWER> </ANSWER>.\\
\\
Response:\\
<ANSWER> Human</ANSWER>
\end{mybackground}

\begin{mybackground}
\textbf{Explicit Case:} \\
Instruction: \\
Based on the given audio, identify the source of the whoop. Please choose the answer from
the following options: [’Human’, ’Bird’, ’Dog’, ’Machine’].\\
Output the thinking process in <THINK> </THINK> and final answer in <ANSWER>
</ANSWER>.\\
\\
Response:\\
<THINK>\\
Hmm, this whoop sounds pretty human. It’s definitely not a bird, dog, or machine. If it were
a machine, it wouldn’t have that kind of emotional or vocal quality. Let’s see if I’m making
a leap here. If it were a bird, the whoop would likely be more of a chirp or tweet. A dog
might bark or whine, but certainly not whoop. So, if it’s not a machine, it must be a human.
That seems to fit best with what I’m hearing. Am I making a leap here? Let’s see if this
holds up under scrutiny. If it were a human, wouldn’t the whoop be more of an emotional
expression or a social interaction? Yes, that makes sense. So, it’s probably a human whoop.\\
</THINK>\\
<ANSWER>Human</ANSWER>
\end{mybackground}

We confirm that RL training utilizing explicit reasoning prompts induced test-time scaling behavior. This approach significantly increased both the model's response length during the inference stage. For instance, on the MMAU benchmark, the average response length for the Qwen2.5-Omni baseline using direct answering is 11.208 tokens. In contrast, CoT-trained models built upon this same base model generate significantly longer outputs: R1-AQA produces an average reasoning length of 42.269 tokens, while the closed-source SARI model reports about 500 tokens (in its Figure 3).

\section{Benchmark and Framework Details }
\label{sec:app_E}
We primarily evaluate model reasoning performance using accuracy on multiple-choice questions, and further conduct reasoning analysis under the CAFE framework. The main evaluations and metrics used are: 
\paragraph{MMAU Benchmark}\cite{sakshi2024mmau} We evaluate the model on the MMAU test-mini split, which features demanding audio QA tasks requiring advanced reasoning. Model performance is reported as multiple-choice accuracy. The MMAU-v05.15.25 release revises approximately 25\% of the questions and answers from the original MMAU benchmark to improve clarity, accuracy, and overall quality. In addition, around 5\% of the audio files are refined to enhance acoustic consistency and signal fidelity. 

The official MMAU benchmark had not been updated during our initial experiments, particularly during the development of the MPAR$^2$ framework with different reward functions. As a result, all experiments were conducted using the original version of MMAU. To ensure experimental consistency, all results reported in the main paper are based on this previous version.

After the release of MMAU-v05.15.25, we further evaluated the trained models on the updated benchmark to assess their performance under the latest setting and to enable direct comparison with current state-of-the-art models.

\paragraph{MMAR Benchmark}~\cite{ma2025mmar} The benchmark is designed to evaluate the reasoning capabilities of Audio Language Models in realistic acoustic environments. It consists of 1,000 audio–question–answer samples collected from open-domain videos and curated through expert annotation and quality control. Unlike prior benchmarks that focus on isolated audio categories, MMAR includes mixed and overlapping audio compositions. For fair comparison with the MMAU dataset, which reports results only on isolated categories, Tables~\ref{table:benchmark_res} and~\ref{table:benchmark_3B} present MMAR results under isolated audio settings. Queries are further organized into four hierarchical reasoning levels, ranging from low-level acoustic analysis to high-level semantic and culturally grounded inference.

\paragraph{CAFE Framework} As detailed in Section \ref{sec:cafe}, CAFE quantifies the perception and utilization of audio events during reasoning. Leveraging ground-truth audio and QA pairs, it evaluates performance via metrics covering perception accuracy, reasoning accuracy, etc.
Notably, in our event extraction design (Section \ref{sec:CAFE_overiew}), we established a specific "Neutral" category. For audio events that are present but unnecessary for answering the question (i.e., "NOT in 'required events'"), the framework explicitly instructs the evaluator: "Do NOT count these as errors". This rule applies provided the model does not incorrectly use them for logical deduction. Therefore, CAFE evaluates whether the model captures necessary information and avoids misusing irrelevant details. It does not mechanically require the model to list all voices.

\section{Supplementary Results}
\label{sec:s_res}
\subsection{MMAU (v05.15.25) Results}
As shown in Table \ref{tab:mmau_table}, when evaluated on MMAU-v05.15.25, the MPAR$^2$ model achieves a substantial improvement over the baseline Qwen2.5-Omni-7B, which is consistent with the performance trends observed on the original MMAU benchmark. In addition, compared with previous models, MPAR$^2$ delivers highly competitive results.
\label{mmau_new}
\begin{table}[t]
    \centering
    \small 
    \resizebox{1\linewidth}{!}{
    \begin{tabular}{l r r r r}
        \toprule
        Model & Sound & Music & Speech & Avg \\ 
        \midrule        
        \rowcolor{gray!20}
        \multicolumn{5}{c}{\textbf{Large Audio Language Models}} \\
        Omni-R1 & 81.70 & 73.40 & 76.00 & 77.00 \\
        MiMo-Audio & 81.68 & \textbf{74.25} & 68.17 & 74.70 \\
        Qwen2.5-Omni-7B & 78.10 & 65.90 & 70.60 & 71.50 \\
        Kimi-Audio & 75.68 & 66.77 & 62.16 & 68.20 \\
        Qwen2-Audio-Instruct & 67.27 & 56.29 & 55.26 & 59.60 \\
        \midrule
        \rowcolor{gray!20}
        \multicolumn{5}{c}{\textbf{Large Audio Reasoning Models}} \\
        Audio-Thinker & \underline{81.98} & \textbf{74.25} & \textbf{76.88} & \textbf{77.70} \\
        Step-Audio 2 & \textbf{84.04} & 73.56 & 75.15 & \underline{77.58} \\
        Audio Flamingo 3 & 79.58 & \underline{73.95} & 66.37 & 73.30 \\
        Audio-Reasoner & 67.87 & 69.16 & 66.07 & 67.70 \\
        \midrule
        \rowcolor{gray!20}
        \multicolumn{5}{c}{\textbf{Commercial Models}} \\
        Gemini 2.5 Flash & 73.27 & 65.57 & \underline{76.58} & 71.80 \\
        GPT-4o Audio & 64.56 & 56.29 & 66.67 & 62.50 \\
        \midrule
        \textbf{MPAR$^2$-7B} & 79.70 & 73.30 & 75.60 & 76.20 \\ 
        \bottomrule
    \end{tabular}
     }
    \caption{Performance Comparison on MMAU-v05.15.25 (Test-mini). Results for other methods are sourced from the MMAU Leaderboard: MMAU-v05.15.25.
    The best results are highlighted in \textbf{bold}, and the second-best scores are \underline{underlined}.}
    \label{tab:mmau_table}
    \vspace{-10pt}
\end{table}

\subsection{Analysis of Audio-Focused Attention}
\label{sec:attention}
\begin{figure}[t]
    \centering
    \scalebox{1}{\includegraphics[width=\columnwidth]{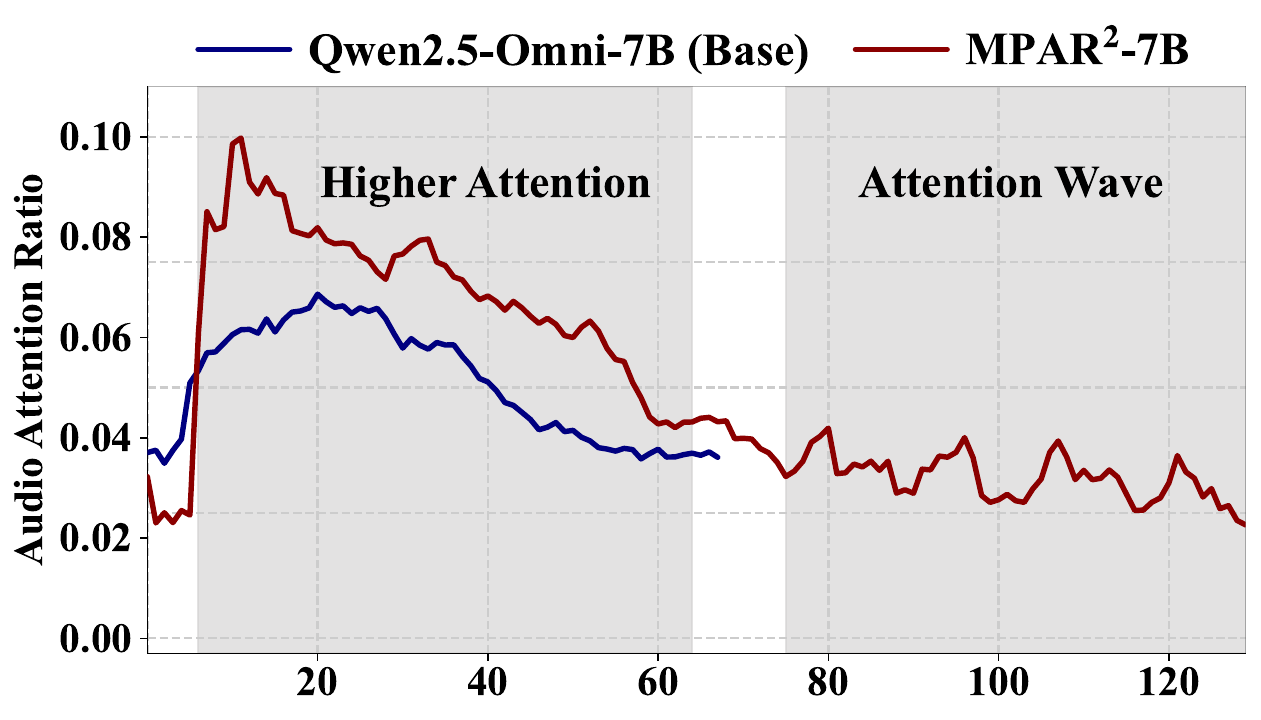}}
    \caption{Audio attention ratio across reasoning token length.}
    \label{fig:Attention}
    \vspace{-10pt}
\end{figure}

We further investigate how MPAR$^2$ training reshapes the model’s attention distribution over audio tokens.
Specifically, we calculate the attention ratios between generated tokens and input audio. 
As illustrated in Figure \ref{fig:Attention}, we visualize the attention dynamics by plotting the attention ratios of the base model and MPAR$^2$ across reasoning token length.
The results indicate that MPAR$^2$ maintains a consistently higher attention level on audio compared to the vanilla Qwen2.5-Omni.
Notably, attention peaks during the early reasoning stage, driven by the initial perceptual processing. Following this peak, rather than decaying, the attention exhibits wave motions, reflecting the model's continuous retrieval with audio context to address perception-related sub-questions.
This pattern suggests that MPAR$^2$ enhances low-level grounding at the input stage while also reinforcing audio-conditioned reasoning in the later decoding process. Together, these results indicate that MPAR$^2$ encourages the model to more effectively integrate acoustic cues throughout the perception–reasoning structure.

\subsection{Discussion of "Audio Perception Decay"}
\label{sec:decay_analysis}
We argue that "Audio Perception Decay" is by no means that "long reasoning is fundamentally harmful in audio reasoning". Instead, we aim to reveal a specific bottleneck: current audio models lack the ability to stably anchor acoustic cues across long generated sequences. Perception decay prevents extended reasoning from delivering its intended positive impact. As the reasoning chain elongates, the model gradually forgets or hallucinates audio events, which degrades the accuracy of subsequent logical deductions.

Meanwhile, we posit that \textit{forcing lengthy reasoning on simple perception tasks is unnecessary}. Our primary analysis supporting the Audio Perception Decay conclusion is based on the high-difficulty, reasoning-intensive MMAR benchmark, which features mixed overlapping audio compositions and spans four hierarchical levels from basic acoustic analysis to high-level semantic and cultural inference. It is inherently a highly difficult and reasoning-heavy benchmark.

Furthermore, on both MMAR and the perception-oriented MMAU, our MPAR$^2$
method significantly improves both perception and overall reasoning performance. This suggests that we do not consider long reasoning universally inefficient. Rather, by enhancing perception during reasoning, we successfully unlock the true potential of CoT in LALMs.

\subsection{\texorpdfstring{MPAR$^2$}{MPAR2}-3B Results}
\label{sec_c_3b}
\begin{table*}[t]
\centering
\tiny 
\setlength{\defaultaddspace}{1pt}
\renewcommand{\arraystretch}{0.9}
\resizebox{\linewidth}{!}{
\begin{tabular}{@{}lcccccccc@{}}
\toprule
\multirow{2}{*}{\textbf{Model}} & 
\multicolumn{4}{c}{\textbf{MMAU (Test-mini)}} & 
\multicolumn{4}{c}{\textbf{MMAR}} 
\\ \cmidrule(lr){2-5} \cmidrule(lr){6-9}

& Sound$\uparrow$ & Music$\uparrow$ & Speech$\uparrow$ & Avg$\uparrow$ & Sound$\uparrow$ & Music$\uparrow$ &  Speech$\uparrow$ & Avg$\uparrow$ \\ 
\midrule

Qwen2.5-Omni-3B & 70.27 & 60.48 & 59.16 & 63.30 & \underline{53.94}  & \textbf{46.12} & 53.74 & 53.80 \\
Qwen2.5-Omni-7B & 69.40 & \underline{66.80} & 61.60 & 65.90 & \underline{53.94} & 42.23 & \underline{58.84} & 55.20 \\
\textbf{MPAR$^2$-3B} & \underline{76.42} & 65.11 & \underline{69.04} & \underline{70.17} & 53.68 & 41.56 & 58.45 & \underline{55.62} \\

\textbf{MPAR$^2$-7B} & \textbf{77.98} & \textbf{71.30} & \textbf{74.47} & \textbf{74.59} & \textbf{62.40} & \underline{44.30} & \textbf{64.26} & \textbf{60.32} \\

\bottomrule
\end{tabular}
}
\caption{Accuracy (\%) on MMAU (original) and MMAR benchmark.}
\label{table:benchmark_3B}
\end{table*}

\paragraph{Performance Enhancement Trajectory Remains Consistent at the 3B Scale}
As detailed in Table~\ref{table:benchmark_3B}, MPAR$^2$-3B significantly outperforms its baseline, elevating accuracy on MMAU from 63.30\% to 70.17\% and on MMAR from 53.80\% to 55.62\%. Mirroring the 7B results, widespread gains are evident across Sound, Music, and Speech domains. These findings confirm that the efficacy of MPAR$^2$ is agnostic to model size, effectively empowering smaller models with robust audio perception and reasoning capabilities.

\paragraph{Similar Mitigation of Perception Decay is also Observed at the 3B Scale}
As illustrated in Figure~\ref{fig:3B_analysis} (a)-(b), the MPAR$^2$-3B model successfully mitigates the trend of perception decay, maintaining robust accuracy even as reasoning chains expand. This consistency confirms that our framework effectively stabilizes perceptual fidelity irrespective of model size.

\paragraph{Adaptive Reasoning Length Persists at the 3B Scale}
As illustrated in Figure~\ref{fig:3B_analysis} (c), MPAR$^2$-3B effectively aligns its reasoning length with task complexity, generating notably longer chains for the reasoning-intensive MMAR benchmark compared to MMAU. Specifically, the model produces concise paths for simple tasks while reserving expanded trajectories for complex acoustic scenarios. This demonstrates a learned ability to dynamically scale reasoning depth, thereby optimizing the balance between computational efficiency and quality.
\begin{table}[t]
\centering
\resizebox{1\linewidth}{!}{
\setlength{\tabcolsep}{5pt}
\begin{tabular}{@{}lccccc@{}}   
\toprule
\textbf{Model} 
&  \textbf{$\text{Acc}_{\textit{per}}$~$\uparrow$}  
&  \textbf{$\text{Err}_{\textit{per}}$~$\downarrow$}
& \textbf{$\text{Err}_{\textit{use}}$~$\downarrow$}
&  \textbf{$\text{Err}_{\textit{omit}}$~$\downarrow$}
& \textbf{$\text{MMAR}_{\textit{acc}}$~$\uparrow$}\\
\midrule
\addlinespace[0.5mm] 

Qwen2.5-Omni-7B & 31.74  & 29.61 & 28.69 & 57.37 & 55.20 \\
Qwen2.5-Omni-3B & 28.54  & 39.35 & 30.9 & 52.14 & 53.80 \\

\midrule
\textbf{MPAR$^2$-7B} & \textbf{63.51}  & \textbf{23.14} & \textbf{7.74} & \textbf{30.59} & \textbf{60.32} \\

\textbf{MPAR$^2$-3B} & \underline{61.04}  & \underline{23.39} & \underline{18.74} & \underline{32.46} & \underline{55.62} \\
\bottomrule
\end{tabular}
}
\caption{Results of various audio models on the MMAR benchmark within the CAFE framework. All results are averaged over three inference runs.}
\label{table:cafe_3B}
\end{table}

\paragraph{Robust Perceptual Performance in 3B Scale} Table~\ref{table:cafe_3B} reveals that this improvement is scale-invariant: MPAR$^2$-3B achieves a perception accuracy ($\text{Acc}_{\text{per}}$) of 61.04\%, doubling the performance of its baseline (28.54\%) and delivering results comparable to the 7B variant. This highlights the method's potential to significantly enhance perception even in smaller-scale models.

\subsection{Ablation Study of Reasoning Mechanisms}
\label{sec_ablation}

\begin{table}[h]
\centering
\resizebox{\linewidth}{!}{
\large
\setlength{\tabcolsep}{6pt}
\renewcommand{\arraystretch}{1.15}
\begin{tabular}{@{}lcccc@{}}
\toprule
\textbf{Method} 
& \textbf{MMAU $\uparrow$} 
& \textbf{MMAR $\uparrow$}
& \textbf{CAFE Acc$_{per}$ $\uparrow$}
& \textbf{Avg. Tokens} \\
\midrule
\addlinespace[0.5mm]
Base & 65.90 & 55.20 & 31.74 & 9.21 \\
Direct-answer GRPO & \underline{70.64} & \underline{57.20} & \underline{32.54} & 6.29 \\
Plain CoT GRPO& 68.32 & 54.60 & 27.34 & 549.36 \\
\midrule
\textbf{MPAR$^2$-7B} & \textbf{74.59} & \textbf{60.32} & \textbf{63.51} & 497.20 \\
\bottomrule
\end{tabular}
}
\caption{Results of direct-answer GRPO and plain CoT GRPO under the same training settings. Base model is Qwen2.5-Omni-7B. Plain CoT GRPO refers to RL training that only includes format and answer rewards. Token count statistics based on MMAR results}
\label{table:ablation_app}
\end{table}

As shown by the ablation results in Table \ref{table:ablation_app}, direct-answer GRPO consistently outperforms the base, indicating that RL itself is effective for improving audio QA performance. In contrast, plain CoT RL underperforms direct-answer GRPO and yields the lowest perceptual accuracy, suggesting that simply introducing CoT reasoning is not a stable strategy for audio QA and that the problem of perception decay becomes particularly severe under this setting. 
Notably, MPAR$^2$ substantially surpasses plain CoT RL, showing that the limitation does not lie in explicit reasoning itself, but rather in the lack of perception-aware reasoning. 
Furthermore, MPAR$^2$ also outperforms direct-answer GRPO, demonstrating that explicit reasoning can be effective when guided by perceptual enhancement.
Meanwhile, compared to plain CoT training, MPAR$^2$ reduces the token budget to some extent.

MPAR² is not intended to claim absolute state-of-the-art performance over every recent audio reasoning model. Its primary goal is to diagnose and mitigate a specific bottleneck: \textit{unstable anchoring of acoustic cues during extended reasoning}. Final benchmark accuracy is jointly influenced by audio perception, inherent logical reasoning ability, and the ability to integrate perceived evidence into multi-step reasoning. Differences in backbone strength, training data, and reward objectives may therefore allow systems such as Audio-Thinker and Step-Audio to achieve higher final accuracy on some reasoning-intensive benchmarks, even when MPAR² achieves better perception accuracy.
\definecolor{ured}{RGB}{228,26,28}
\definecolor{ugreen}{RGB}{77,175,74}
\usetikzlibrary{decorations.markings}
\definecolor{ugreen}{RGB}{098,190,166}
\definecolor{uyellow}{RGB}{253,186,107}
\definecolor{ured}{RGB}{235,096,070}
\definecolor{upurple}{RGB}{175,135,220} 
\definecolor{ublue}{RGB}{076,135,220}
\definecolor{ugray}{RGB}{150,156,165}
\definecolor{ubrown}{RGB}{176,125,086}

\pgfmathsetlengthmacro{\BarW}{12pt}
\pgfmathsetlengthmacro{\BarStep}{16pt}
\pgfmathsetlengthmacro{\GapClusters}{-10pt}

\pgfmathsetlengthmacro{\ShiftEasy}{-2.0*\BarStep - 0.5*\GapClusters}
\pgfmathsetlengthmacro{\ShiftMedium}{ -1.0*\BarStep - 0.5*\GapClusters}
\pgfmathsetlengthmacro{\ShiftSynth}{-2.5*\BarStep}
\pgfmathsetlengthmacro{\ShiftHard}{0.0*\BarStep - 0.5*\GapClusters}
\pgfmathsetlengthmacro{\ShiftIE}{1.0*\BarStep - 0.5*\GapClusters}
\pgfmathsetlengthmacro{\ShiftReason}{2.0*\BarStep - 0.5*\GapClusters}

\pgfmathsetlengthmacro{\ShiftSingle}{ -1.5*\BarStep + 0.5*\GapClusters}
\pgfmathsetlengthmacro{\ShiftMixture}{-0.5*\BarStep + 0.5*\GapClusters}
\pgfmathsetlengthmacro{\ShiftPercep}{  0.5*\BarStep + 0.5*\GapClusters}
\pgfmathsetlengthmacro{\ShiftHigh}{    1.5*\BarStep + 0.5*\GapClusters}

\tikzset{
  hatchbar/.style={
    postaction={draw=none, pattern=north east lines}
  },
  hatchCross/.style={
    postaction={draw=none, pattern=crosshatch}
  }
}

\begin{figure*}[tb]
\centering

\begin{minipage}[t]{0.32\textwidth}
\centering
\begin{tikzpicture}
\begin{axis}[
    width=\linewidth,
    height=0.75\linewidth,
    ymajorgrids,
    grid style=dashed,
    xmin=-1, xmax=19,
    ymin=0, ymax=1,
    xtick={0,2,...,18},
    xticklabels={25,35,45,55,65,75,85,95,105,115},
    ytick={0,0.2,0.4,0.6,0.8,1},
    ticklabel style={font=\tiny},
    label style={font=\scriptsize},
    legend style={
        font=\tiny,
        fill opacity=0.5,
        draw=none,
        text opacity=1,
        at={(axis description cs:0.98,0.98)},
        anchor=north east
    }
]
\addplot[ured, mark=o, mark size=1.6pt,line width=1pt] coordinates {
   (0,0.503676) (2,0.531789) (4,0.446075)
   (6,0.480567) (8,0.534387)
   (10,0.362482) (12,0.331319)
   (14,0.323611) (16,0.220476) (18,0)
};
\addplot[ugreen, mark=square,mark size=1.6pt, line width=1pt] coordinates {
   (0,0.550000) (2,0.422383) (4,0.487462)
   (6,0.293413) (8,0.333333)
   (10,0.212121) (12,0.366667)
   (14,0.200000) (16,0.133333) (18,0.0000)
};
\legend{Perception Accuracy, Reasoning Accuracy}
\end{axis}
\end{tikzpicture}

\vspace{-0.4em}
{\small \hspace*{1.4em}(a) Qwen2.5-Omni-3B (Base)}
\end{minipage}\hfill%
\begin{minipage}[t]{0.32\textwidth}
\centering
\hspace*{-2cm}
\begin{tikzpicture}
\begin{axis}[
    width=\linewidth,
    height=0.75\linewidth,
    ymajorgrids,
    grid style=dashed,
    xmin=-1, xmax=19,
    ymin=0, ymax=1,
    xtick={0,2,...,18},
    xticklabels={224,272,320,368,416,464,512,560,608,656},
    ytick={0,0.2,0.4,0.6,0.8,1},
    ticklabel style={font=\tiny},
    label style={font=\scriptsize},
    legend style={
        font=\tiny,
        fill opacity=0.5,
        draw=none,
        text opacity=1,
        at={(axis description cs:0.98,0.98)},
        anchor=north east
    }
]
\addplot[ured, mark=o, mark size=1.6pt,line width=1pt] coordinates {
   (0,0.433333) (2,0.527652) (4,0.488034)
   (6,0.461382) (8,0.449098)
   (10,0.423418) (12,0.590714)
   (14,0.422151) (16,0.433636) (18,0.534014)
};
\addplot[ugreen, mark=square,mark size=1.6pt, line width=1pt] coordinates {
   (0,0.433333) (2,0.372727) (4,0.407692)
   (6,0.465854) (8,0.433766)
   (10,0.412500) (12,0.373333)
   (14,0.403704) (16,0.30000) (18,0.414286)
};
\legend{Perception Accuracy, Reasoning Accuracy}
\end{axis}
\end{tikzpicture}

\vspace{-0.4em}
{\small \hspace*{-1.5cm}(b) MPAR$^2$-3B}
\end{minipage}\hfill%
\begin{minipage}[t]{0.32\textwidth}
\hspace*{-2.cm}
\centering
\begin{tikzpicture}
\begin{axis}[
  ybar,
  bar width=8pt,
  width=1.3\linewidth,
  height=0.75\linewidth,
  ymin=360, ymax=680,
  ytick={400, 440, 480, 520, 560, 600, 640},
  symbolic x coords={MMAU (Test-mini),MMAR},
  xtick={MMAU (Test-mini),MMAR},
  enlarge x limits=0.52,
  axis line style={thick},
  tick style={thick,draw=none},
  label style={font=\normalsize},
  yticklabel style={font=\scriptsize},
  xticklabel style={font=\scriptsize,yshift=6pt},
  ymajorgrids,
  grid style={dashed, gray!30},
  legend style={
    font=\tiny,
    draw=none,
    fill=none,
    text opacity=1,
    rounded corners,
    at={(0.0,1.56)},
    anchor=north west,
    legend cell align=left,
    nodes={anchor=mid west},
    /tikz/every node/.append style={yshift=6pt}
  },
  legend image code/.code={
    \path[#1, draw=none, fill opacity=0.95] (0pt,-1pt) rectangle (8pt,7pt);
  },
  legend columns=3,
  transpose legend, 
  clip=false,
  nodes near coords,
  nodes near coords align={vertical},
  every node near coord/.append style={
    font=\tiny,
    text=black,
    yshift=2pt,
  },
  point meta=y,
  nodes near coords={\pgfmathprintnumber[fixed,precision=2]{\pgfplotspointmeta}},
]
\addplot+[forget plot, draw=uyellow!80, fill=uyellow!40, bar shift=\ShiftEasy+5pt]
  coordinates {(MMAU (Test-mini),398)};

\addplot+[forget plot, draw=uyellow!80, fill=uyellow!70, bar shift=\ShiftMedium]
  coordinates {(MMAU (Test-mini),445)};

\addplot+[forget plot, draw=uyellow!80, fill=uyellow!90, bar shift=\ShiftHard-5pt]
  coordinates {(MMAU (Test-mini),486)};

\addplot+[forget plot, draw=ublue!80, fill=ublue!50, bar shift=\ShiftIE-10pt]
  coordinates {(MMAU (Test-mini),382)};

\addplot+[forget plot, draw=ublue!80, fill=ublue!80, bar shift=\ShiftReason-15pt]
  coordinates {(MMAU (Test-mini),443)};

\addplot+[forget plot, draw=ugreen!60, fill=ugreen!30, bar shift=\ShiftSingle+16pt]
  coordinates {(MMAR,545)};

\addplot+[forget plot, draw=ugreen!60, fill=ugreen!50, bar shift=\ShiftMixture+11pt]
  coordinates {(MMAR,623)};

\addplot+[forget plot, draw=ured!80, fill=ured!50, bar shift=\ShiftPercep+6pt]
  coordinates {(MMAR,552)};

\addplot+[forget plot, draw=ured!80, fill=ured!70, bar shift=\ShiftHigh+1pt]
  coordinates {(MMAR,633)};

\addlegendimage{ybar, draw=uyellow!80, fill=uyellow!40}
\addlegendentry{Easy}

\addlegendimage{ybar, draw=uyellow!80, fill=uyellow!70}
\addlegendentry{Medium}

\addlegendimage{ybar, draw=uyellow!80, fill=uyellow!90}
\addlegendentry{Hard}

\addlegendimage{ybar, draw=ublue!80, fill=ublue!50}
\addlegendentry{Information Extraction}

\addlegendimage{ybar, draw=ublue!80, fill=ublue!80}
\addlegendentry{Reasoning}

\addlegendimage{ybar, draw=ugreen!80, fill=ugreen!30}
\addlegendentry{Single-Source}

\addlegendimage{ybar, draw=ugreen!80, fill=ugreen!50 }
\addlegendentry{Acoustic Mixtures}

\addlegendimage{ybar, draw=ured!80, fill=ured!50}
\addlegendentry{Perception}

\addlegendimage{ybar, draw=ured!80, fill=ured!70}
\addlegendentry{High-level Cognitive}
    
\draw[thick]
  (rel axis cs:0,0) -- (rel axis cs:1,0);

\end{axis}
\end{tikzpicture}

\vspace{-0.4em}
{\small \hspace*{-2cm}(c) Reasoning Token Length of\\ \hspace*{-2cm} MPAR$^2$-3B on Benchmarks}
\end{minipage}

\caption{
(a)-(b) show the accuracy of MPAR$^2$-3B vs. Base across token lengths. (c) is the Reasoning token length of MPAR$^2$-3B across different
task categories on the MMAU and MMAR benchmark.
}
\label{fig:3B_analysis}
\end{figure*}

\section{Training Details}
\label{sec:app_C}

\subsection{SFT and RL Setting}
\label{sec:traing_setting}
\begin{mybackground}
SFT:\\
Epochs: 3\\
Train dataset: $Data_{SFT}$\\
Total batch size: 16 \\
Learning rate: 5e-5 \\

GRPO:\\
Train type: LoRA \\
LoRA rank: 8 \\
LoRA alpha: 32 \\
Torch dtype: bfloat16 \\
Max completion length: 2048 \\
Per device train batch size: 1 \\
Per device eval batch size: 1 \\
Gradient accumulation steps: 2 \\
Learning rate: 1e-6 \\
Warmup ratio: 0.05 \\
Num generations: 8 \\
Temperature: 1.0 \\
Top p: 0.99 \\
Top k: 50 \\
Deepspeed: ZeRO-3 
\end{mybackground}

\subsection{Data Production Process}
\begin{figure*}[t]
    \centering
    \includegraphics[width=1\textwidth]{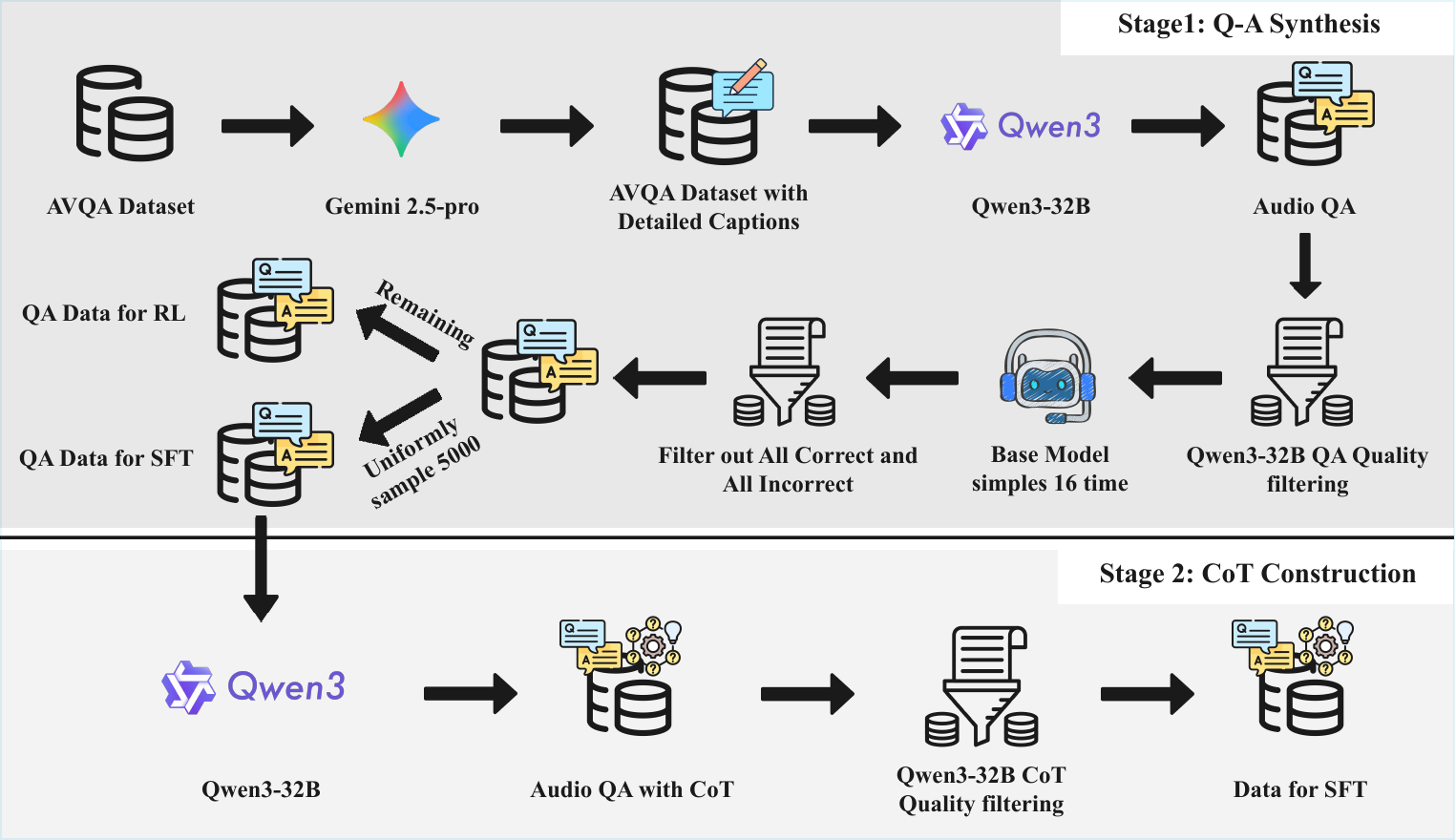}    
    \caption{
     The overview of the data construction pipeline.
    }
    \label{fig:data}
\end{figure*}

The overview of the data construction is in Figure \ref{fig:data}.
To teach LALMs this structured reasoning pattern and prepare data for RL, we draw from
the AVQA dataset~\cite{yang2022avqa}, which is designed for video question answering. 
We first employed Gemini-2.5-Pro to generate detailed captions for the AVQA dataset. Based on these captions, Qwen3-32B synthesized reasoning-oriented QA pairs across four dimensions: timbre characteristics, temporal patterns, pitch contours, and rhythmic structures. We then filtered the QA pairs for logic and consistency, and assess difficulty by running the initial base model 16 times, discarding samples that were completely correct or incorrect, resulting in 46,544 QA samples. From the filtered data pool, we curated a balanced set of 5,000 samples across four aspects and different time durations, while the remaining 41,544 samples $Data_{RL}$ were reserved for Stage-2 RL. Finally, we prompt to use these 5,000 samples to generate CoT reasoning paths by Qwen3-32B. Following CoT quality filtering, 4,600 instances, denoted as $Data_{SFT}$, were selected for Stage-1 supervised fine-tuning. 

Below are the prompts used in the production process.

\subsubsection{Gemini-2.5-Pro Caption Prompt}
\begin{mybackground}
You are an expert audio event analyst. Your task is to carefully analyze the entire audio and extract all meaningful events as comprehensively as possible, including source entities, time intervals and captions.
\\
For each extracted event, you must:

1. Identify the possible sound sources within the event and make a brief descriptions for it.

2. Provide the accurate time interval [start time,end time] for each event when it occurs within the audio.

3. Provide a detailed, precise, and contextually rich caption of what happens in that event interval.

Do not infer or imagine information that cannot be clearly heard. The max number of extracted events is 5.\\



Please separate different parts using the following format:\\
\textless event1\textgreater\\
\quad \textless description\textgreater...\textless/description\textgreater\\
\quad \textless time\textgreater[start time, end time]\textless/time\textgreater\\
\quad \textless caption\textgreater...\textless/caption\textgreater\\
\textless/event1\textgreater\\
\textless event2\textgreater\\
\quad \textless description\textgreater...\textless/description\textgreater\\
\quad \textless time\textgreater[start time, end time]\textless/time\textgreater\\
\quad \textless caption\textgreater...\textless/caption\textgreater\\
\textless/event2\textgreater\\
\textless event3\textgreater\\
\quad \textless description\textgreater...\textless/description\textgreater\\
\quad \textless time\textgreater[start time, end time]\textless/time\textgreater\\
\quad \textless caption\textgreater...\textless/caption\textgreater\\
\textless/event3\textgreater\\
......\\

Here is a sample. Please strictly follow the format in the sample.\\
\textless event1\textgreater\\
\quad \textless description\textgreater dogs barking \textless/description\textgreater\\
\quad \textless time\textgreater[0.00s, 2.35s]\textless/time\textgreater\\
\quad \textless caption\textgreater the audio captures a brief, intense episode of multiple dogs barking aggressively in an outdoor setting. \textless/caption\textgreater\\
\textless/event1\textgreater\\
\textless event2\textgreater\\
\quad \textless description\textgreater...\textless/description\textgreater\\
\quad \textless time\textgreater[start time, end time]\textless/time\textgreater\\
\quad \textless caption\textgreater...\textless/caption\textgreater\\
\textless/event2\textgreater\\
\textless event3\textgreater\\
\quad \textless description\textgreater...\textless/description\textgreater\\
\quad \textless time\textgreater[start time, end time]\textless/time\textgreater\\
\quad \textless caption\textgreater...\textless/caption\textgreater\\
\textless/event3\textgreater......\\

Please extract events from the follow audio: 

\end{mybackground}

\subsubsection{QA Generate Prompt}
Event Counting QA Generate Prompt:
\begin{mybackground}
We are constructing training data to enhance audio perception and reasoning in audio large language models (LLMs).

Your job is to create a single, high-quality multiple-choice question that tests whether a model can perform numerical reasoning and quantitative analysis by \textbf{listening} to a complete audio clip.\\

\textbf{Input Audio Events (Chronological Order)}: {events\_description}\\

\textbf{Task Description:}

You are provided with the ground truth sequence of sound events above.
Your goal is to generate ONE multiple-choice question (MCQ) focusing on \textbf{Event Counting} \& \textbf{Numerical Reasoning}.\\

\textbf{Step 1: Suitability Check (CRITICAL)}

First, analyze the event list provided above. Ask yourself:

1. Are there distinct, countable discrete events (e.g., 'bark', 'gunshot', 'footstep') rather than just continuous ambience (e.g., 'wind', 'static', 'silence')?

2. Is there enough variety or repetition to form a valid numerical question (e.g., counting total occurrences, comparing counts of two different sources)?

If the content consists mainly of continuous noise, ambiguous sounds, or a single non-repeatable event that makes counting trivial or impossible, \textbf{you must output exactly}:

Not suitable for this hallucination type.\\

\textbf{Step 2: Question Generation (If Suitable)}

If the audio events support numerical reasoning, generate ONE MCQ.

The question must target the \textbf{auditory experience}. Do NOT ask about the text descriptions directly.
The question must be answerable by listening to the audio and counting/analyzing the sounds.\\

\textbf{Focus Areas:}

- \textbf{Total Count}: 'How many times is the [specific sound] heard?'

- \textbf{Source Comparison}: 'Did the dog bark more times than the cat meowed?'

- \textbf{ Sequence Logic}: 'After the first door slam, how many footsteps follow?'

(Note: Since you do not have exact timestamps, focus on the count and order of events, not the speed or rate per minute.)\\

\textbf{STRICT CONSTRAINT:}

The question must NOT contain phrases like 'According to the list', 'In the description', or 'text'.
It must sound like a natural question asked to someone who has just listened to the recording.\\

\textbf{Output Format (if suitable):}

Question: \textless question about the sound itself\textgreater

A. \textless option A\textgreater

B. \textless option B\textgreater

C. \textless option C\textgreater

D. \textless option D\textgreater

Correct answer: \textless the correct option letter\textgreater
\end{mybackground}

Case Study of Event Counting
\begin{mybackground}
\textbf{Question}: How many distinct cawing events are clearly heard before the audio becomes chaotic with overlapping caws?\\
\textbf{Choice}: A. One   B. Two   C. Three   D. Four \\
\textbf{Correct answer}: C
\end{mybackground}

Pitch Contour QA Generate Prompt
\begin{mybackground}
We are constructing training data to enhance audio perception and reasoning in audio large language models (LLMs). 

Your job is to create a single, high-quality multiple-choice question that tests whether a model can perform deep inference based on the \textbf{dynamic pitch contour (frequency modulation)} by \textbf{listening} to a complete audio clip.\\

\textbf{Input Audio Events (Chronological Order):} {events\_description}\\

\textbf{Task Description:}

You are provided with the ground truth sequence of sound events above. 

Your goal is to generate ONE multiple-choice question (MCQ) focusing on \textbf{Pitch Contour \& Semantic Inference}.\\

\textbf{Step 1: Suitability Check (CRITICAL)}

Analyze the event list provided above. To support a question about pitch contour, the audio must contain sources with detectable tonal properties or modulation. Ask yourself:

1. Does the audio contain \textbf{speech} (where intonation conveys meaning like sarcasm or questions)?

2. Does it contain \textbf{tonal machinery or vehicles} (where speed/movement affects pitch, e.g., Doppler effect, acceleration)?

3. Does it contain \textbf{musical or biological sounds} (e.g., bird calls, singing) where frequency changes involve meaning?

\textbf{If the audio consists ONLY of unpitched sounds} (e.g., 'rain', 'static noise', 'footsteps', 'door slam', 'wind') where pitch analysis is irrelevant or impossible, \textbf{you must output exactly:}

Not suitable for this hallucination type.\\

\textbf{Step 2: Question Generation (If Suitable)}

If the audio contains tonal or pitch-varying events, generate ONE MCQ.

The question must require the model to hear the curve of the sound (rising, falling, wavering) and deduce the underlying cause, intent, or movement.\\

\textbf{Focus Areas (Audio-Centric):}

- \textbf{Intonation \& Intent:} 'The speaker's pitch rises sharply at the end. What does this suggest about their certainty?' (Reasoning: Statement vs. Question)

- \textbf{Doppler Effect (Physics):} 'As the [vehicle] sound gets louder and then fades, the pitch drops significantly. What does this confirm about its movement?' (Reasoning: Passing the listener)

- \textbf{Mechanical State:} 'The pitch of the engine continuously increases without dropping. What does this imply about the machine's operation?' (Reasoning: Acceleration/Revving)

- \textbf{Emotional State:} 'The tremor or wavering pitch in the voice suggests what emotion?' (Reasoning: Fear/Excitement\\

\textbf{STRICT CONSTRAINT:}

The question must NOT contain phrases like 'According to the list', 'In the description', or 'text'. 

It must sound like a natural question asked to someone who has just closed their eyes and listened to the recording.\\

\textbf{Output Format (if suitable):}

Question: \textless question about the pitch/tone/intonation\textgreater

A. \textless option A\textgreater

B. \textless option B\textgreater

C. \textless option C\textgreater

D. \textless option D\textgreater

Correct answer: \textless the correct option letter\textgreater
\end{mybackground}

Case Study of Pitch Contour

\begin{mybackground}
\textbf{Question}:  A siren is heard starting low, rising quickly to a high pitch, and then holding that pitch for a while before stopping. What is this pattern most likely indicating?\\
\textbf{Choice}:  A. A fire truck arriving at the scene   B. An emergency evacuation alert   C. A vehicle accelerating away   D. A malfunctioning car alarm \\
\textbf{Correct answer}: B
\end{mybackground}

Rhythmic Structure QA Generate Prompt
\begin{mybackground}
We are constructing training data to enhance audio perception and reasoning in audio large language models (LLMs).

Your job is to create a single, high-quality multiple-choice question that tests whether a model can perform deep inference based on the \textbf{rhythmic structure and temporal regularity} by \textbf{listening} to a complete audio clip.\\

\textbf{Input Audio Events (Chronological Order):} {events\_description}\\

\textbf{Task Description:}

You are provided with the ground truth sequence of sound events above. 
Your goal is to generate ONE multiple-choice question (MCQ) focusing on \textbf{Rhythmic Structure \& Behavioral Inference}.\\

\textbf{Step 1: Suitability Check (CRITICAL)}

Analyze the event list provided above. To support a question about rhythm, the audio must contain sounds that repeat or form a pattern. Ask yourself:

1. Does the audio contain \textbf{repetitive impulsive sounds} (e.g., 'footsteps', 'typing', 'heartbeat', 'clapping', 'knocking')?

2. Does it contain \textbf{rhythmic machinery or engines} (e.g., 'train wheels', 'clock ticking', 'idling engine')?

3. Does it contain \textbf{music or percussion}?

\textbf{If the audio consists ONLY of continuous amorphous noise} (e.g., 'wind', 'water flow', 'static') \textbf{OR single isolated events} (e.g., 'one gunshot', 'a single drop', 'one scream') where no rhythm exists, \textbf{you must output exactly:}

Not suitable for this hallucination type.\\

\textbf{Step 2: Question Generation (If Suitable)}

If the audio contains rhythmic patterns, generate ONE MCQ.

The question must require the model to hear the pattern of the sound (regularity, tempo, acceleration, chaos) and deduce the underlying behavior or state.\\

\textbf{Focus Areas (Audio-Centric):}

- \textbf{Activity State (Tempo):} 'The footsteps transition from a slow, steady walking pace to a rapid, frantic rhythm. What does this suggest about the subject?' (Reasoning: Leisure vs. Fleeing)

- \textbf{Stability (Regularity):} 'The engine sound changes from a steady hum to an irregular, sputtering pattern. What does this indicate?' (Reasoning: Mechanical failure)

- \textbf{Coordination (Synchronization):} 'The clapping sounds represent a large group hitting the beat in perfect unison. What context does this imply?' (Reasoning: Organized audience/Performance vs. Chaotic crowd)

- \textbf{Environmental Context:} 'The rhythm of the typing is sporadic and hesitant, with long pauses. What does this suggest about the typist?' (Reasoning: Thinking/Uncertainty vs. Professional transcription)\\

\textbf{STRICT CONSTRAINT}:

The question must NOT contain phrases like 'According to the list', 'In the description', or 'text'. 
It must sound like a natural question asked to someone who has just closed their eyes and listened to the recording.\\

\textbf{Output Format (if suitable):}

Question: \textless question about the pitch/tone/intonation\textgreater

A. \textless option A\textgreater

B. \textless option B\textgreater

C. \textless option C\textgreater

D. \textless option D\textgreater

Correct answer: \textless the correct option letter\textgreater
\end{mybackground}

Case Study of Rhythmic Structure
\begin{mybackground}
\textbf{Question}: The pig’s vocalizations occur in a tightly spaced, repetitive sequence at the start of the audio. What does this rhythmic pattern most likely suggest about the pig’s emotional or behavioral state?  A. It is calmly and slowly exploring its environment. \\
\textbf{Choice}: A. It is calmly and slowly exploring its environment.   B. It is experiencing distress or agitation.   C. It is communicating with a distant group of pigs.   D. It is playing or engaging in a harmless interaction with another animal.   \\
\textbf{Correct answer}: B
\end{mybackground}

Temporal Event Reasoning QA Generate Prompt
\begin{mybackground}
We are constructing training data to enhance audio perception and reasoning in audio large language models (LLMs). 

Your job is to create a single, high-quality multiple-choice question that tests whether a model can perform \textbf{Temporal Logic and Causal Reasoning} by \textbf{listening} to a complete audio clip.\\

\textbf{Input Audio Events (Chronological Order):} {events\_description}\\

\textbf{Task Description:}

You are provided with the ground truth sequence of sound events above. 
Your goal is to generate ONE multiple-choice question (MCQ) focusing on \textbf{Sequence, Causality, or Temporal Relationship}.\\

\textbf{Step 1: Suitability Check (CRITICAL)}

Analyze the event list provided above. To support temporal or causal reasoning, the audio must contain a \textbf{sequence of multiple distinct events}. Ask yourself:

1. Are there \textbf{at least two distinct events} happening in succession (e.g., 'thunder' then 'rain', 'footsteps' then 'door open')?

2. Is there a logical link (cause-and-effect) or a clear chronological order to test?

\textbf{If the audio consists of a SINGLE event} (e.g., just 'dog barking') OR only continuous background noise (e.g., 'static', 'city traffic') where no sequence exists, \textbf{you must output exactly:}

Not suitable for this hallucination type.\\

\textbf{Step 2: Question Generation (If Suitable)}

If the audio contains a valid sequence of events, generate ONE MCQ.

The question must require the model to \textbf{track the order of sounds} or \textbf{deduce the cause} of a sound based on what happened before/after.

(Note: Do NOT ask about specific duration in seconds, as you do not have exact timestamps. Focus on relative order and logic.)\\

\textbf{Focus Areas (Audio-Centric):}

- \textbf{Chronological Order:} 'Which sound occurred immediately after the glass shattered?' (Reasoning: Tracking sequence).

- \textbf{Causality/Reaction:} 'The sudden braking sound was immediately followed by a crash and shouting. What does this sequence imply?' (Reasoning: Cause and Effect).

- \textbf{Interruption:} 'The music was abruptly cut off by which specific sound?' (Reasoning: Identifying the disruptor).

- \textbf{Reverse Inference:} 'The sound of the audience clapping suggests that what event likely just finished?' (Reasoning: Inferring the preceding context).\\

\textbf{STRICT CONSTRAINT:}

The question must NOT contain phrases like 'According to the list', 'In the description', or 'text'. 
It must sound like a natural question asked to someone who has just closed their eyes and listened to the recording.\\

\textbf{Output Format (if suitable):}

Question: \textless question about the pitch/tone/intonation\textgreater

A. \textless option A\textgreater

B. \textless option B\textgreater

C. \textless option C\textgreater

D. \textless option D\textgreater

Correct answer: \textless the correct option letter\textgreater
\end{mybackground}

Case Study of Etemporal logic
\begin{mybackground}
\textbf{Question}: A single dog begins barking, and soon after, multiple dogs join in. Which of the following sounds is most likely to have triggered the initial barking? \\
\textbf{Choice}: A. The low-frequency electronic tone   B. The person sighing or groaning   C. The sound of footsteps   D. The rustling of clothing  \\
\textbf{Correct answer}: A
\end{mybackground}

Timbre Characteristics QA Generate Prompt
\begin{mybackground}
We are constructing training data to enhance audio perception and reasoning in audio large language models (LLMs). 
Your job is to create a single, high-quality multiple-choice question that tests whether a model can perform \textbf{Temporal Logic and Causal Reasoning} by \textbf{listening} to a complete audio clip.\\

\textbf{Input Audio Events (Chronological Order):} {events\_description}\\

\textbf{Task Description:}

You are provided with the ground truth sequence of sound events above. 
Your goal is to generate ONE multiple-choice question (MCQ) focusing on \textbf{Sequence, Causality, or Temporal Relationship}.\\

\textbf{Step 1: Suitability Check (CRITICAL)}

Analyze the event list provided above. To support temporal or causal reasoning, the audio must contain a \textbf{sequence of multiple distinct events}. Ask yourself:

1. Are there \textbf{at least two distinct events} happening in succession (e.g., 'thunder' then 'rain', 'footsteps' then 'door open')?

2. Is there a logical link (cause-and-effect) or a clear chronological order to test?

\textbf{If the audio consists of a SINGLE event} (e.g., just 'dog barking') \textbf{OR only continuous background noise} (e.g., 'static', 'city traffic') where no sequence exists, \textbf{you must output exactly:}

Not suitable for this hallucination type.\\

\textbf{Step 2: Question Generation (If Suitable)}

If the audio contains a valid sequence of events, generate ONE MCQ.

The question must require the model to \textbf{track the order of sounds} or \textbf{deduce the cause} of a sound based on what happened before/after.

(Note: Do NOT ask about specific duration in seconds, as you do not have exact timestamps. Focus on relative order and logic.)\\

\textbf{Focus Areas (Audio-Centric):}

- \textbf{Chronological Order:} 'Which sound occurred immediately after the glass shattered?' (Reasoning: Tracking sequence).

- \textbf{Causality/Reaction:} 'The sudden braking sound was immediately followed by a crash and shouting. What does this sequence imply?' (Reasoning: Cause and Effect).

- \textbf{Interruption:} 'The music was abruptly cut off by which specific sound?' (Reasoning: Identifying the disruptor).

- \textbf{Reverse Inference:} 'The sound of the audience clapping suggests that what event likely just finished?' (Reasoning: Inferring the preceding context).\\

\textbf{STRICT CONSTRAINT:}

The question must NOT contain phrases like 'According to the list', 'In the description', or 'text'. 

It must sound like a natural question asked to someone who has just closed their eyes and listened to the recording.\\

\textbf{Output Format (if suitable):}

Question: \textless question about the pitch/tone/intonation\textgreater

A. \textless option A\textgreater

B. \textless option B\textgreater

C. \textless option C\textgreater

D. \textless option D\textgreater

Correct answer: \textless the correct option letter\textgreater
\end{mybackground}

Case Study of Timbre Characteristics
\begin{mybackground}
\textbf{Question}: Based on the sound characteristics of the buzzing in the background, what is the most likely physical source producing this continuous, low-frequency noise? \\
\textbf{Choice}:  A. A swarm of bees inside a wooden beehive   B. A malfunctioning electrical device emitting a hum   C. A distant, low-flying airplane   D. A person humming a tune consistently and softly  \\
\textbf{Correct answer}:  A
\end{mybackground}

\subsection{QA Filter Prompt}
\begin{mybackground}
\textbf{Role}

You are an expert Data Quality Evaluator for Audio-Text Reasoning Datasets. Your goal is to strictly filter out low-quality QA pairs and keep only those that require genuine logical reasoning.\\

\textbf{Objective}

Evaluate the following Audio QA pair. You must determine if the Question requires logical deduction, causal inference, or temporal analysis of the events described in the Caption.\\

\textbf{Input Data}

- \textbf{Audio Caption}: {caption}

- \textbf{Question and Answer}: {question}\\

\textbf{Evaluation Criteria} (The ReasoningLitmus Test)\\

\textbf{1. POSITIVE INDICATORS} (High Score / KEEP)

The question requires the model to:

\textbf{Infer Cause/Effect}: Why did the sound stop?(Requires understanding the preceding event).

\textbf{Analyze Sequence (Temporal)}: What happened immediately after the explosion?

\textbf{Deduce State/Intent}: Based on the footsteps and breathing, is the person running or walking?

\textbf{Synthesize Multiple Clues}: Combining background noise + specific actions to determine the location.\\

\textbf{2. NEGATIVE INDICATORS} (Low Score / DISCARD)

\textbf{Simple Pattern Matching}: The answer is just a word lifted directly from the caption (e.g., Caption says a red car, Question asks what color is the car?).

\textbf{Common Sense / General Knowledge}: Can be answered without the audio caption (e.g., Do birds fly?).

\textbf{Summarization: Describe the audio}(This is generation, not reasoning).

\textbf{Unsolvable/Hallucination}: The answer assumes facts not present in the caption.\\

\textbf{Task}

1. Analyze the relationship between the Caption, Question, and Answer.

2. Assign a \textbf{Score (1-5)} based on reasoning depth.

- \textbf{5}: Complex reasoning (multi-hop, causal, or temporal).

- \textbf{4}: Clear deduction required.

- \textbf{3}: Simple inference.

- \textbf{2}: Direct text retrieval / keyword matching.

- \textbf{1}: Common sense or irrelevant.

3. Make a final \textbf{Decision}: KEEP(Score \textgreater= 4) or DISCARD(Score \textless4).\\

\textbf{Output Format}

Return ONLY a valid JSON object:

{{

analysis: Brief explanation of the reasoning logic required.,

score: \textless int\textgreater,

decision: KEEPor DISCARD

}}
\end{mybackground}

\subsection{\texorpdfstring{MPAR$^2$}{MPAR2}CoT Generate Prompt}
\begin{mybackground}
\textbf{Role:}

You are an advanced AI assistant specializing in Audio Reasoning and Chain-of-Thought (CoT) synthesis. You also act as a strict auditor to ensure data quality.\\

\textbf{Task:}

Your task is to synthesize the provided input data into a structured CoT format using specific XML tags, and then perform a \textbf{critical validation (Review)} of the reasoning process.

You will be given:

1.\textbf{Question:} The main query about the audio.

2.\textbf{Final Answer:} The ground truth answer

3.\textbf{Caption:} A time-interval detailed description of the audio events (Ground Truth)

4.\textbf{Model Outputs:} A list of step-by-step sub-questions and answers generated logically in previous steps.\\

\textbf{Output Format Requirements:}

You must output the content strictly inside the following XML structure:

\textless thinking\textgreater

\textless perception\textgreater

1. [start\_time, end\_time]: Description of event A.

2. [start\_time, end\_time]: Description of event B.

... (List ALL events from the Caption, chronologically)

\textless /perception\textgreater

\textless reasoning\textgreater

1. Sub-question: [First step from Model Outputs]

 Answer: [Answer to first step]
 
... (Include all steps from the Model Outputs)

\textless /reasoning\textgreater

\textless review\textgreater

1. Evidence Check: [Simulate a \textbf{Re-listening}process. Verify if the audio events cited in the 'Reasoning' are factually supported by the events listed in 'Perception'. Check for hallucinations, misinterpretations, or missing details. \textbf{Note: Treat the 'Perception' content as the audio itself; do not refer to 'captions', 'text', or 'provided descriptions'.}]

2. Logic Check: [Evaluate the logical validity. Does the conclusion naturally follow from the evidence? Is the overall chain coherent?]

\textless /review\textgreater

\textless /thinking\textgreater

\textless answer\textgreater

[The Final Answer provided in the input]

\textless /answer\textgreater\\

\textbf{Directives:}

1.\textbf{Perception Section (Full Extraction):}
Translate the \textbf{ENTIRE} provided `Caption` into a structured, chronological list of time intervals and event descriptions.
\textbf{Do not filter, summarize, or omit} any events, even if they seem irrelevant to the specific Question.
Format: `[t1, t2]: content`.

2.\textbf{Reasoning Section:}
Directly utilize the provided `Model Outputs`. Organize them into a numbered list of Sub-questionand Answerpairs.
Ensure the reasoning logic flows smoothly.

3.\textbf{Review Section (Critical Audit):}

\textbf{Evidence Check:} Perform a \textbf{Re-Perception} check.

\textbf{Context:} Imagine you are re-checking the audio stream directly.

\textbf{Validity:} Did the reasoning cite sounds that actually exist in the `Perception` list?

\textbf{Accuracy:} Did the reasoning interpret the sound properties correctly?

\textbf{Constraint:} \textbf{Strictly avoid} phrases like according to the captionor the text says. Instead, use phrases like The audio contains..., I hear..., or The event at [timestamp] shows....

\textbf{Logic Check:} Verify the soundness of the deductive process. Ensure there are no logical leaps or contradictions.

4.\textbf{Answer Section:} State the final answer clearly.\\
\textbf{Input Data:}\\
\textbf{Question:}

{ORIGINAL\_QUESTION}\\
\textbf{Final Answer:}

{FINAL\_ANSWER}\\
\textbf{Caption:}

{caption\_w\_time}\\
\textbf{Model Outputs (Reasoning Chain):}
{sub\_question\_list\_generated}

\end{mybackground}
\subsection{CoT Data Filter Prompt}
\begin{mybackground}
\textbf{Role}

You are an expert evaluator for Audio-Language Models. Your task is to audit a Chain-of-Thought(CoT) process generated by an AI model. You will assess how well the model hearsthe audio (Perception), thinksabout it (Reasoning), and auditsits own conclusion (Review).\\
\textbf{Input Data}

1. \textbf{Original Question}: The user's query.

2. \textbf{Final Answer}: The ground truth answer.

3. \textbf{Caption}: Ground truth events with precise time ranges.

4. \textbf{CoT}: The step-by-step reasoning process containing \textless perception\textgreater, \textless reasoning\textgreater, and \textless review\textgreater tags.\\

\textbf{Evaluation Dimensions}

\textbf{Dimension 1: Perception Evaluation (The Ear)}

Analyze specific claims about audio events/timestamps in the \textless perception\textgreater and \textless reasoning\textgreater sections.

Evaluate based on:

1.\textbf{Accuracy}: Does the event described match the Caption?

2.\textbf{Hallucination/Omission}: Are there invented sounds or missed critical sounds?

\textbf{Dimension 2: Reasoning Evaluation (The Brain)}

Analyze the logical flow in the \textless reasoning\textgreater section.

Evaluate each step on:

1.\textbf{Utility}: Is this step necessary for solving the Original Question?

2.\textbf{Factuality}: Is the statement factually true based on the audio content?

3.\textbf{Logical Validity}: Does the conclusion naturally follow from the cited evidence?

\textbf{Dimension 3: Review Evaluation (The Auditor)}

Analyze the \textless review\textgreater section, specifically the Evidence Checkand Logic Check.

Evaluate based on:

1.\textbf{Evidence Re-verification}: 

Did the model correctly re-examine and re-perceive the audio events cited in the reasoning? 

Did it accurately confirm whether the events exist in the perception data? 

\textbf{Critical}: Did it successfully identify valid evidence versus hallucinated evidence?

2.\textbf{Rationality Check}: 
Did the model correctly assess the logical coherence of the entire chain?
Did it ensure that the Final Answer is the only logical conclusion derived from the evidence?\\

\textbf{Scoring \& Output Format}

Output only two numbers (0-10) strictly in accordance with the following format:
Reasoning\_Score/Review\_Score

\textbf{Scoring Criteria:}

\textbf{Reasoning\_Score (num1)}: Rate the quality of Dimension 1 (Perception) and Dimension 2 (Reasoning). 10 = Perfect audio detection and flawless logic.

\textbf{Review\_Score (num2)}: Rate the quality of Dimension 3 (Review). 10 = The model performed a rigorous, accurate self-audit that correctly validated the evidence and logic. 0 = The review was superficial, inaccurate, or failed to catch obvious errors.\\

\textbf{Input Data:}\\

\textbf{Question:}
{ORIGINAL\_QUESTION}\\

\textbf{Final Answer:}
{FINAL\_ANSWER}\\

\textbf{Caption:}
{caption\_w\_time}\\

\textbf{Model Outputs (Reasoning Chain):}
{sub\_question\_list\_generated}

\end{mybackground}

\section{Reward prompt}
\label{sec:app_D}
\subsection{Perception Score Prompt}
\begin{mybackground}
You are an expert audio perception evaluator. I will give you a record containing:

1. A Detailed Audio Caption (Ground Truth): A comprehensive, factual text description of the audio events.

2. A Question and its Correct Answer: To determine which audio events are 'critical' for the task.

3. A Model Perception Output: The content within the \texttt{\textless perception\textgreater} tags generated by the model, describing events with timestamps.

Your task is to evaluate the fidelity, precision, and completeness of the Model Perception against the Ground Truth, and output a single numeric score from {0, 0.1, 0.2, ..., 1.0}. 
You must output only the score with no explanation or extra text.\\

Evaluate based on the following CRITICAL principles:

1. Audio Hallucination (Strict) — The model must NOT report events that do not exist in the 'Detailed Audio Caption'. 

Reporting a sound that is completely absent (e.g., hearing a siren when the description only mentions birds) is a fatal failure.

2. Content Accuracy \& Sequential Logic — While evaluating the model's generated timestamps, focus on:

- Event Identity: Does the model correctly identify the sound sources described in the Ground Truth? (e.g., distinguishing 'footsteps' from 'knocking').

- Chronological Flow: Does the sequence of events in the model's output match the narrative order of the Ground Truth? (e.g., if the description says 'a door opens then slams', the model must not place the slam before the opening).

3. Critical Event Coverage (Relevance) — The model must capture all 'Key Events' necessary to answer the provided 'Question'. 
Compare with the Question/Answer pairs: if the answer depends on a specific sound cue, omitting this specific event in the \textless perception\textgreater phase is a critical failure.

4. Consistency \& Identity — The model should describe the same audio source consistently across different timestamps (unless the sound evolves). 
Avoid contradictory descriptions for the same ongoing event.

5. Redundancy \& Conciseness — The perception output should be dense and informative. 
Penalize distinct 'loops' (repeating the exact same phrase for adjacent timestamps) or extreme verbosity that adds no new details.\\

Scoring guideline

1.0 = Flawless. Perfectly matches the Ground Truth description. Events are correctly identified and listed in the correct logical order. Concise.

0.8-0.9 = Excellent. Accurate detection of all key events described. The sequence is logical. Maybe minor verbosity.

0.5-0.7 = Mediocre. The KEY event was detected, but the description is vague, or information irrelevant to the question has been omitted.

0.2-0.4 = Poor. Misses a KEY event needed for the Answer, or misidentifies a sound source. Sequence is disorderly compared to the description.

0.0-0.1 = Severe. HALLUCINATION (inventing sounds not in the description), or total failure to identify the main audio event.\\

Penalty guideline (Apply these cumulatively to reduce the score):

\textbf{[CRITICAL PENALTY] (Set Score to 0.0 $\sim$ 0.2)} : 

- Hallucinating an event not present in the Ground Truth description.

- Misidentifying the main sound source (e.g., 'gunshot' vs 'drum').

- Missing the specific audio cue required to answer the Question.\\
\textbf{[MODERATE PENALTY] (-0.3 $\sim$ -0.5)} : 

- Sequential Logic Error (Events are listed in an order contradicting the description).

- Significant omission of details mentioned in the description.\\
\textbf{[MINOR PENALTY] (-0.1 $\sim$ -0.2)} :

- Excessive wordiness or repetitive phrasing without new information.

- Vague descriptions (e.g., 'noise' instead of 'dog barking') if the Ground Truth is specific.\\

Operational rule: Always output only one score (0-1 in 0.1 increments).
Now evaluate the following record and output only the score.

The Detailed Audio Caption (Ground Truth) is:{caption\_text}

The Question is:{question\_text}

The Correct Answer is:{answer\_text}

The Model perception to evaluate is:{cot\_text}

\end{mybackground}
\subsection{Step-Level Reasoning Score Prompt}
\begin{mybackground}
You are an expert logic and reasoning evaluator for Audio-LLMs. I will give you a record containing: 

1. A Detailed Audio Caption (Model Perception): A comprehensive text description of the audio events.

2. A User Question and Context: The goal of the reasoning.

3. A Reasoning History: The steps taken so far.

4. The CURRENT STEP: The specific sub-question or reasoning step to evaluate now.\\

Your task is to evaluate the \textbf{validity, necessity, and audio-grounding} of the CURRENT STEP only, and output a single numeric score from {0, 0.1, 0.2, ..., 1.0}. 
You must output only the score with no explanation or extra text

Evaluate based on the following Micro-Level dimensions:\\

--- CRITERIA: Local Quality Check ---

1. Usefulness: Is this specific reasoning step useful for answering the main question?

2. Evidence-Based Conclusion: Is the content of this step supported by the provided 'Audio Caption'?

- Every claim must align with specific events, sound sources, or acoustic details described in the caption.

3. Criticality \& Efficiency: Is this step a logical next move based on the [Reasoning History]?

- Penalize 'tangential reasoning' (analyzing irrelevant noise) or redundant repetition of previous steps.\\

Scoring guideline: 

1.0 = Perfect. The step is firmly grounded in the caption, necessary, and logically follows the history.

0.8-0.9 = Strong. Good step, but maybe slightly inefficient or the evidence citation is slightly vague.

0.5-0.7 = Mediocre. Relevant, but weak grounding (making assumptions not explicitly in the caption). Logic holds but is messy.

0.2-0.4 = Weak. The step makes a claim not supported by the caption, or merely repeats previous steps without adding value.

0.0-0.1 = Failed. Completely incoherent, visual hallucination, or factual contradiction with the caption (e.g., claiming a sound exists when caption implies silence).\\

Penalty guideline (Apply these cumulatively to reduce score):

\textbf{[CRITICAL PENALTY] (Set Score to 0.0 $\sim$ 0.1): }

- 'Factual Contradiction': The step claims a specific sound or event occurs which is explicitly absent or contradicted by the Caption.\\

Operational rule: Always output only one score (0-1). If [Current Step] is empty, return 0.0.\\

The Detailed Audio Caption is:{caption\_text}

The User Question is:{question\_text}

The Reasoning History is:{history\_text}

The CURRENT STEP to evaluate is:{current\_step\_text}

\end{mybackground}
\subsection{Holistic Level Reasoning Score Prompt}
\begin{mybackground}
You are an expert logic and reasoning evaluator for Audio-LLMs. I will give you a record containing: 

1. A Detailed Audio Caption (Model Perception): A comprehensive text description of the audio events.

2. A Question and its Correct Answer.

3. The COMPLETE Model Reasoning: The entire chain of thought generated by the model.\\

Your task is to evaluate the \textbf{logical architecture, coherence, efficiency, and final derivability} of the entire process, and output a single numeric score from {0, 0.1, 0.2, ..., 1.0}. 
You must output only the score with no explanation or extra text.\\

Evaluate based on the following Macro-Level dimensions:\\

--- CRITERIA: Holistic Logical Architecture ---

1. Goal-Orientation: Is the reasoning path linear and directed towards the [Correct Answer]? 

   - Penalize circular logic.
   
2. Causal Dependency: Does Step B legitimately follow Step A? 

   - Penalize 'Logic Jumps' where a conclusion appears out of nowhere without a preceding premise defined in the audio caption.
   
3. Error Propagation Check: Does an early error render the rest of the chain invalid?

   - If Step 1 is wrong (e.g., misidentifying a gender or sound source compared to the caption), and subsequent steps rely on it, the whole chain collapses.
   
4. Final Derivability: Does the reasoning naturally flow to the [Correct Answer]?

   - The conclusion must be the inevitable result of the reasoning steps, not a sudden guess.
   
5. Efficiency \& Conciseness: Is the length of the reasoning proportional to the complexity of the question?

   - \textbf{Penalize 'Over-Analysis'}: If the question is simple (e.g., 'Is there a dog?'), the reasoning should be short. Writing a 500-word essay for a simple question is a failure.
   
   - \textbf{Penalize Repetition}: Check if the model repeats the same analysis in different words just to make the chain longer.\\

Scoring guideline: 

1.0 = Perfect. Every step is necessary, concise, and the logic flows flawlessly from the caption evidence to the correct conclusion.

0.8-0.9 = Strong. Good logic, but maybe slightly verbose or includes one unnecessary step, yet the path is valid.

0.5-0.7 = Mediocre. Logic holds but is \textbf{bloated} or unfocused. Contains repetitive analysis or over-explains simple facts found in the caption.

0.2-0.4 = Weak. Major Logic Jumps, or the reasoning is \textbf{excessively long} and tedious without adding value (Filibustering).

0.0-0.1 = Failed. The reasoning contradicts the final answer, relies on 'Fatal Error Propagation', or is complete nonsense.\\

Penalty guideline (Apply these cumulatively):

 \textbf{[CRITICAL PENALTY] (Set Score to 0.0 $\sim$ 0.2): }

   - 'Fatal Error Propagation': Early false premise (contradicting the Audio Caption) corrupts the entire remaining chain.
   
   - 'Contradiction': The reasoning concludes something different from the actual Correct Answer provided.\\
 \textbf{[MODERATE PENALTY] (-0.2 $\sim$ -0.4)} : 

   - 'Bloated Reasoning': The reasoning is too long for the problem's difficulty (e.g., 10 steps for a Yes/No question).
   
   - 'Irrelevance': Wasting steps on analyzing audio events that are present in the caption but do not help answer the specific Question.\\

Operational rule: Always output only one score (0-1). \\

The Detailed Audio Caption is:{caption\_text}

The Question is:{question\_text}

The Correct Answer is:{answer\_text}

The COMPLETE Model Reasoning is:{full\_reasoning}

\end{mybackground}
\subsection{Review Score Prompt}
\begin{mybackground}
You are a critical meta-evaluator for the "Self-Correction" (Review) phase of an Audio-LLM. 

Your task is to judge whether the [Review Content] effectively audits, verifies, and corrects the [Model Reasoning].\\

Input Data:

1. [Detailed Audio Caption] (Model Perception): {caption\_text}

2. [Ground Truth Annotations] (Fact Reference): {ground\_truth\_text}

3. [User Question]: {question\_text}

4. [Correct Answer] (Ground Truth): {answer\_text}

5. [Model Reasoning] (Target to Audit): {reasoning\_text}

6. [Review Content] (The Audit Output): {review\_text}

Task: Output a single score {{0.0, 0.1, ..., 1.0}} for the [Review Content].\\

Evaluation Criteria (Review Quality):

1. Evidence Verification (Content Alignment):

- Does the Review explicitly verify that every event cited in the [Model Reasoning] actually exists in the [Detailed Audio Caption]?

- Did it catch "Hallucinations" where the Reasoning cites a sound (e.g., "dog barking") that is completely absent from the Caption?

- \textbf{CRITERIA}: The Review must confirm that the \textbf{evidence} used in reasoning is physically present in the text description.

2. Temporal \& Causal Logic Audit:

- Does the Review check the narrative sequence? (e.g., if Reasoning says "A causes B", did the Review check if the Caption describes A happening before or leading into B?)

- Did it catch chronological errors where the Reasoning flips the order of events described in the text?

3. Logical Integrity Check:
- Does the Review ensure the conclusion is strictly derived from the perceived evidence?
- Did it flag any "Logic Jumps" (conclusions without premises) in the Reasoning?

4. Genuine Error Correction \& Answer Verification (Anti-Rubber-Stamping):
- \textbf{ANSWER CHECK (CRITICAL)}: Compare the conclusion/answer in [Model Reasoning] with the [Correct Answer]. 

    - If Reasoning leads to a WRONG answer, the Review \textbf{MUST} detect and flag this failure.
    
    - If Reasoning leads to a CORRECT answer, the Review must validate the logic flow.
    
- \textbf{HALLUCINATION CHECK}: If [Model Reasoning] contains factual errors (contradictions with Caption), the Review \textbf{MUST} point them out.

- \textbf{PENALTY RULES}:
    - \textbf{Score 0.0 IMMEDIATELY} if the [Model Reasoning] concludes with a wrong answer (mismatch with [Correct Answer]) but the Review states "The reasoning is correct" or "The answer is valid".
    - \textbf{Score 0.0 IMMEDIATELY} if the Review approves reasoning that contradicts the Audio Caption (e.g., approving a sound that isn't in the description).\\

Scoring Guide:

- 1.0: Perfect Audit. The review rigorously checked evidence presence, sequential logic, AND accurately validated the final answer against Ground Truth.

- 0.8-0.9: Strong. Good check, caught major issues, but maybe missed a minor detail in the description.

- 0.5-0.7: Generic Validation. Says "Logic is good" without citing specific text evidence. (Rubber-stamping).

- 0.2-0.4: Weak. Fails to catch obvious hallucinations or logic jumps.

- 0.0-0.1: [FATAL] The Review acts as a "Yes-Man" (Rubber-Stamp) for an INCORRECT Answer. It approves a reasoning path that leads to a result different from the [Correct Answer].

Output Rule: Return ONLY the numeric score.

\end{mybackground}

\section{AI Tools}
We used GPT-5.5 solely for writing refinement. 
All research ideas, experimental design, experiments, and analyses were conducted by the authors.

\end{document}